\documentclass[10pt,a4paper,twocolumn]{scrartcl}

\usepackage{amsmath}
\usepackage{amssymb}

\usepackage{graphicx}

\usepackage{cite}

\usepackage{color} 

\usepackage[labelfont=bf,font=footnotesize,labelsep=period,justification=justified,format=plain]{caption}

\makeatletter
\renewcommand{\@biblabel}[1]{\quad#1.}
\makeatother

\date{}

\usepackage[normalem]{ulem}
\newcommand{\m}{m}
\newcommand{\B}{B}
\newcommand{\ns}{C}
\newcommand{\x}{x}
\newcommand{\y}{y}
\newcommand{\bx}{\beta_\x}
\newcommand{\by}{\beta_\y}
\newcommand{\beh}[1]{\beta_{#1}}
\newcommand{\s}{s}
\newcommand{\ac}{a}
\newcommand{\X}{X}
\newcommand{\Y}{Y}
\newcommand{\mo}{N_\text{tot}}
\newcommand{\tref}{\mu}
\newcommand{\refeq}[1]{Eq.~\textbf{\ref{ecn:#1}}}
\newcommand{\refig}[2]{Fig.~\ref{fig:#1}\emph{#2}}
\newcommand{\meth}{\emph{Methods}}

\newcommand{\suppdata}{\emph{Protocol S1} and \emph{Protocol S2}}
\newcommand{\supptext}{\emph{Text S1}}

\newcommand{\ecnSXVI}{\textbf{S16}}

\newcommand{\comment}[1]{}
\newcommand{\ecngeneral}{3}
\newcommand{\ecnindep}{6}
\newcommand{\ecnkbehav}{7}
\newcommand{\M}{M}

\title{Collective Animal Behavior from Bayesian Estimation and Probability Matching}
\author{Alfonso P\'erez-Escudero\footnote{Instituto Cajal, Consejo Superior de Investigaciones Cien\-t\'i\-fi\-cas, Av. Doctor Arce, 37, 28002 Madrid, Spain. Department of Theoretical Physics and Ins\-ti\-tu\-to ``Nicol\'as Cabrera" de F\'isica de Materiales, Universidad Aut\'onoma de Madrid, 28049 Madrid, Spain. E-mail: alfonso.perez.escudero@cajal.csic.es, gonzalo.polavieja@cajal.csic.es}, Gonzalo G. de Polavieja$^{\ast}$}
\begin{document}
\maketitle


\section*{Abstract}
Animals living in groups make movement decisions that depend, among other factors, on social interactions with other group members. Our present understanding of social rules in animal collectives is mainly based on empirical fits to observations, with less emphasis in obtaining first-principles approaches that allow their derivation. Here we show that patterns of collective decisions can be derived from the basic ability of animals to make probabilistic estimations in the presence of uncertainty. We build a decision-making model with two stages: Bayesian estimation and probabilistic matching.
In the first stage, each animal makes a Bayesian estimation of which behavior is best to perform taking into account personal information about the environment and social information collected by observing the behaviors of other animals. In the probability matching stage, each animal chooses a behavior with a probability equal to the Bayesian-estimated probability that this behavior is the most appropriate one. This model derives very simple rules of interaction in animal collectives that depend only on two types of reliability parameters, one that each animal assigns to the other animals and another given by the quality of the non-social information. We test our model by obtaining theoretically a rich set of observed collective patterns of decisions in three-spined sticklebacks, \emph{Gasterosteus aculeatus}, a shoaling fish species. The quantitative link shown between probabilistic estimation and collective rules of behavior allows a better contact with other fields such as foraging, mate selection, neurobiology and psychology, and gives predictions for experiments directly testing the relationship between estimation and collective behavior.

\section*{Author Summary}
Animals need to act on uncertain data and with limited cognitive abilities to survive. It is well known that our sensory and sensorimotor processing uses probabilistic estimation as a means to counteract these limitations. Indeed, the way animals learn, forage or select mates is well explained by probabilistic estimation. Social animals have an interesting new opportunity since the behavior of other members of the group provides a continuous flow of indirect information about the environment. This information can be used to improve their estimations of environmental factors. Here we show that this simple idea can derive basic interaction rules that animals use for decisions in social contexts. In particular, we show that the patterns of choice of \emph{Gasterosteus aculeatus} correspond very well to probabilistic estimation using the social information. The link found between estimation and collective behavior should help to design experiments of collective behavior testing for the importance of estimation as a basic property of how brains work. 

\section*{Introduction}

Animals need to make decisions without certainty in which option is best. This uncertainty is due to the ambiguity of sensory data but also to limited processing capabilities, and is an intrinsic and general property of the representation that animals can build about the world. A general way to make decisions in uncertain situations is to make probabilistic estimations \cite{Box1973,Jaynes2003}. There is evidence that animals use probabilistic estimations, for example in the early stages of sensory perception \cite{Helmholtz1925,Mach1980,Knill2004,Jacobs1999,Knill2003a,Ernst2002,Battaglia2003,Alais2004,Gold2001}, sensory-motor transformations \cite{Kording2004,Kording2006,Gold2007}, learning \cite{Courville2006,Kruschke2006,Tenenbaum2011} and behaviors in an ecological context such as strategies for food patch exploitation \cite{OATEN1977,Biernaskie2009,ALONSO1995} and mate selection \cite{McNamara2006}, among others \cite{Kording2006,J.Valone2006,McNamara2006,Tenenbaum2011}. 

An additional source of information about the environment may come from the behavior of other animals (social information) \cite{Valone2002,Blanchet2010,Dall2005,Giraldeau2002,Wagner2010,King2007}. This information can have different degrees of ambiguity. In particular cases, the behavior of conspecifics directly reveals environmental characteristics (for example, food encountered by another individual informs about the quality of a food patch). Cases in which social information correlates well with the environmental characteristic of interest have been very well studied \cite{Valone1989,Templeton1995,Templeton1996,Smith1999,CLARK1986,Doligez2002,Boulinier1997,Coolen2003,VanBergen2004}. But in most cases social information is ambiguous and potentially misleading \cite{Giraldeau2002,Rieucau2009}. In spite of this ambiguity, there is evidence that in some cases such as predator avoidance \cite{Lima1995,Proctor2001} and mate choice \cite{Nordell1998}, animals use this kind of information.

Social animals have a continuous flow of information about the environment coming from the behaviours of other animals. It is therefore possible that social animals use it at all times, making probabilistic estimations to counteract its ambiguity. If this is the case, estimation of the environment using both non-social and social information might be a major determinant of the structure of animal collectives. In order to test this hypothesis, we have developed a Bayesian decision-making model that includes both personal and social information, that naturally weights them according to their reliability in order to get a better estimate of the environment. All members of the group can then use these improved estimations to make better decisions, and collective patterns of decisions then emerge from these individuals interacting through their perceptual systems.

We show that this model derives social rules that economically explain detailed experiments of decision-making in animal groups \cite{Ward2008,Sumpter2008}. This approach should complement the empirical approach used in the study of animal groups \cite{Ward2008,Sumpter2008,Couzin2003,Sumpter2006,Couzin2005,Katz2011}, finding which mathematical functions should correspond to each experimental problem and to propose experiments relating estimation and collective motion. The Bayesian structure of our model also builds a bridge between the field of collective behavior and other fields of animal behavior, such as optimal foraging theory \cite{OATEN1977,Biernaskie2009,ALONSO1995,J.Valone2006,McNamara2006} and others \cite{J.Valone2006,McNamara2006}. Further, it explicitly includes in a natural way different cognitive abilities, making more direct contact with neurobiology and psychology \cite{Helmholtz1925,Mach1980,Knill2004,Jacobs1999,Knill2003a,Ernst2002,Battaglia2003,Alais2004,Tenenbaum2011}.

\section*{Results}
\subsection*{Estimation model}
We derived a model in which each individual decides from an estimation of which behavior is best to perform. These behaviors can be to go to one of several different places, to choose among some behaviors like forage, explore or run away, or any other set of options. For clarity, here we particularize to the case of choosing the best of two spatial locations, $\x$ and $\y$ (see \supptext\ for more than two options). `Best' may correspond to the safest, the one with highest food density or most interesting for any other reasons.  We assume that each decision maker uses in the estimation of the best location both non-social and social information. Non-social information may include sensory information about the environment (i.e. shelter properties, potential predators, food items), memory of previous experiences and internal states. Social information consists of the behaviors performed by other decision-makers. Each individual estimates the probability that each location, say $\y$, is the best one, using its non-social information ($\ns$) and the behavior of the other individuals ($\B$),
\begin{equation}\label{ecn:1}
P(\Y|\ns,\B),
\end{equation}
where $\Y$ stands for '$\y$ is the best location'. $P(\X|\ns,\B)=1-P(\Y|\ns,\B)$, because there are only two locations to choose from. We can compute the probability in \refeq{1} using Bayes' theorem,
\begin{multline}\label{ecn:bayes}
P(\Y|\ns,\B)=\\
\frac{P(\B|\Y,\ns)P(\Y|\ns)}{P(\B|\X,\ns)P(\X|\ns)+P(\B|\Y,\ns)P(\Y|\ns)}.
\end{multline}
By simply dividing numerator and denominator by the numerator we find an interesting structure,
\begin{equation}\label{ecn:bayesfrac}
P(\Y|\ns,\B)=\frac{1}{1+\ac\, S},
\end{equation}
where
\begin{equation}\label{ecn:a0}
\ac=\frac{P(\X|\ns)}{P(\Y|\ns)}
\end{equation}
and
\begin{equation}\label{ecn:S}
S=\frac{P(\B|\X,\ns)}{P(\B|\Y,\ns)}.
\end{equation}
Note that $a$ does not contain any social information so it can be understood as the``non-social term'' of the estimation. We can also understand $S$ as the ``social term'' because it contains all the social information, although is also depends on the non-social information $\ns$. The non-social term $\ac$ is the likelihood ratio for the two options given only the non-social information. This kind of likelihood ratio is the basis of Bayesian decision-making in the absence of social information \cite{Knill2004,Gold2001,Gold2007,Kording2004,Kording2006}. \refeq{bayesfrac} now tells us that this well known term interacts with the social term $S$ simply through multiplication.


We are seeking a model based on probabilistic estimation that can simultaneously give us insight into social decision-making and fit experimental data. For this reason we simplify the model by assuming that the focal individual does not make use of the correlations among the behaviour of others, but instead assumes their behaviours to be independent of each other. This is a strong hypothesis but allows us to derive simple explicit expressions with important insights. The section `Model including dependencies' at the end of Results shows that this assumption gives a very good approximation to a more complete model that takes into account these correlations. 

The assumption of independence translates in that the probability of a given set of behaviors is just the product of the probabilities of the individual behaviors. We apply it to the probabilities needed to compute $S$ in \refeq{S}, getting
\begin{equation}\label{ecn:behavindep}
P(\B|\Y,\ns)= Z \prod_{i=1}^N P(b_i|\Y,\ns),
\end{equation}
where $\B$ is the set of all the behaviors of the other $N$ animals at the time the focal individual chooses, $\B=\left\{b_{i}\right\}_{i=1}^{N}$, and $b_i$ denotes the behavior of one of them, individual $i$. $Z$ is a combinatorial term counting the number of possible decision sequences that lead to the set of behaviors $B$, that will cancel out in the next step. Substituting \refeq{behavindep} and the corresponding expression for $P(\B|\X,\ns)$ into \refeq{S}, we get
\begin{equation}\label{ecn:Sindep}
S=\prod_{i=1}^N\frac{P(b_i|\X,\ns)}{P(b_i|\Y,\ns)}.
\end{equation}
Instead of an expression in terms of as many behaviors as individuals, it may be more useful to consider a discrete set of behavioral classes. For example, in our two-choice example, these behavioral classes may be `go to $\x$' (denoted $\bx$), `go to $\y$' ($\by$) and `remain undecided' ($\beh{u}$). Frequently, these behavioral classes (or simply `behaviors') will be directly related to the choices, so that each behavior will consist of choosing one option. For example, behaviors $\bx$ and $\by$ are directly related to choices $\x$ and $\y$, respectively. But there may be behaviors not related to any option as the case of indecision, $\beh{u}$, or related to choices in an indirect way. These behaviors can still be informative because they may be more consistent with one of the options being better than the other (for example, indecision may increase when there is a predator, so the presence of undecided individuals may bias the decision against the place where the non-social information suggests the presence of a predator).
Let us consider $L$ different behavioral classes, $\{\beh{k}\}_{k=1}^L$. We do not here consider individual differences for animals performing the same behavior (say, behavior $\beh{1}$), so they have the same probabilities $P(\beh{1}|\X,\ns)$ and $P(\beh{1}|\Y,\ns)$. Thus, if for example the $n_1$ first individuals are performing behavior $\beh{1}$, we have that $\prod_{i=1}^{n_1}\frac{P(b_i|\X,\ns)}{P(b_i|\Y,\ns)}=\left(\frac{P(\beh{1}|\X,\ns)}{P(\beh{1}|\Y,\ns)}\right)^{n_1}$. We can then write \refeq{Sindep} as 
\begin{equation}\label{ecn:Skchoices}
S=\prod_{k=1}^L \s_k^{n_k},
\end{equation}
where $n_k$ is the number of individuals performing behavior $\beh{k}$, and
\begin{equation}\label{ecn:sk}
\s_k=\frac{P(\beh{k}|X,\ns)}{P(\beh{k}|Y,\ns)}.
\end{equation}
The term $\s_k$ is the probability that an individual performs behavior $\beh{k}$ when $x$ is the best option, over the probability that it performs the same behavior when $y$ is the best choice. The higher $\s_k$ the more reliably behavior $\beh{k}$ indicates that $x$ is better than $y$, so we can understand $\s_k$ as the reliability parameter of behavior $\beh{k}$. If $\s_k=\infty$, observing behavior $\beh{k}$ indicates with complete certainty that $x$ is the best option, while for $\s_k=1$ behavior $\beh{k}$ gives no information. For $\s_k<1$, observing behavior $\beh{k}$ favors $y$ as the best option, and more so the closer it is to 0. Note that $P(\beh{k}|\X,\ns)$ and $P(\beh{k}|\Y,\ns)$ are not the actual probabilities of performing behavior $\beh{k}$, but estimates of these probabilities that the deciding animal uses to assess the reliability of the other decision-makers. These estimates may be `hard-wired' as a result of evolutionary adaptation, but may also be subject to change due to learning.

To summarize, using Eqs. \textbf{\ref{ecn:bayesfrac}} and \textbf{\ref{ecn:Skchoices}}, the probability that $\y$ is the best choice, given both social and non-social information is
\begin{equation}\label{ecn:general2choicebayes}
P(\Y|\ns,\B)=\left(1+\ac\prod_{k=1}^L\s_k^{n_k}\right)^{-1},
\end{equation}
with $a$ in \refeq{a0} and $s_k$ in \refeq{sk}.

\subsection*{Decision rule: Probability matching}
We have so far only considered the perceptual stage of decision-making, in which the deciding individual estimates the probability that each behavior is the best one. Now it must decide according to this estimation. A simple decision rule would be to go to $\y$ when $P(\Y|\ns,\B)$ is above a certain threshold. This rule maximizes the amount of correct choices when the probabilities do not change \cite{Neyman1933}, but is not consistent with the experimental data considered in this paper. Applying this deterministic rule strictly, without any noise sources, one would obtain that all individuals behave exactly in the same way when facing the same stimuli, but in the experiments considered here this is not the case. Instead, we used a different decision rule called probability matching, that has been experimentally observed in many species, from insects to humans \cite{Herrnstein1961,Behrend1961,Greggers1993,Kirk1965,Vulkan2000,Wozny2010,Staddon1983}. According to this rule an individual chooses each option with a probability that is equal to the probability that it is the best choice. Therefore, in our case the probability of going to $\y$ ($P_\y$), is the same as the estimated probability that $\y$ is the best location ($P(\Y|\ns,\B)$), so
\begin{equation}\label{ecn:2}
P_\y=P(\Y|\ns,\B).
\end{equation}
Probability matching does not maximize the amount of right choices if we assume that the probabilities stay always the same, but in many circumstances it can be the optimal behavior, such as when there is competition for resources \cite{Fretwell1969,Houston1987}, when the estimated probabilities are expected to change due to learning \cite{Staddon1983,Vulkan2000}, or for other reasons \cite{Vulkan2000,Gaissmaier2008}. 


Finally, using Eqs. \ref{ecn:general2choicebayes} and \ref{ecn:2} we have that the probability that the deciding individual goes to $y$ is
\begin{equation}\label{ecn:general2choice}
P_y=\left(1+\ac\prod_{k=1}^L\s_k^{n_k}\right)^{-1}.
\end{equation}
The assumption of probability matching has the advantage that the final expression for the decision in \refeq{general2choice} is identical to the one given by Bayesian estimation in \refeq{general2choicebayes}, with no extra parameters. Alternative decision rules could be noisy versions of the threshold rule, but at the price of adding at least one extra parameter to describe the noise. Also, decision rules might not depend on estimation alone, but also on other factors or constraints. These more complicated rules fall beyond the scope of this paper.

In the following sections, we particularize \refeq{general2choice} to different experimental settings to test its results against existing rich experimental data sets that have previously been fitted to different mathematical expressions \cite{Ward2008,Sumpter2008}.



\subsection*{Symmetric set-up}
\begin{figure}[!ht]
\begin{center}
\includegraphics[width=\columnwidth]{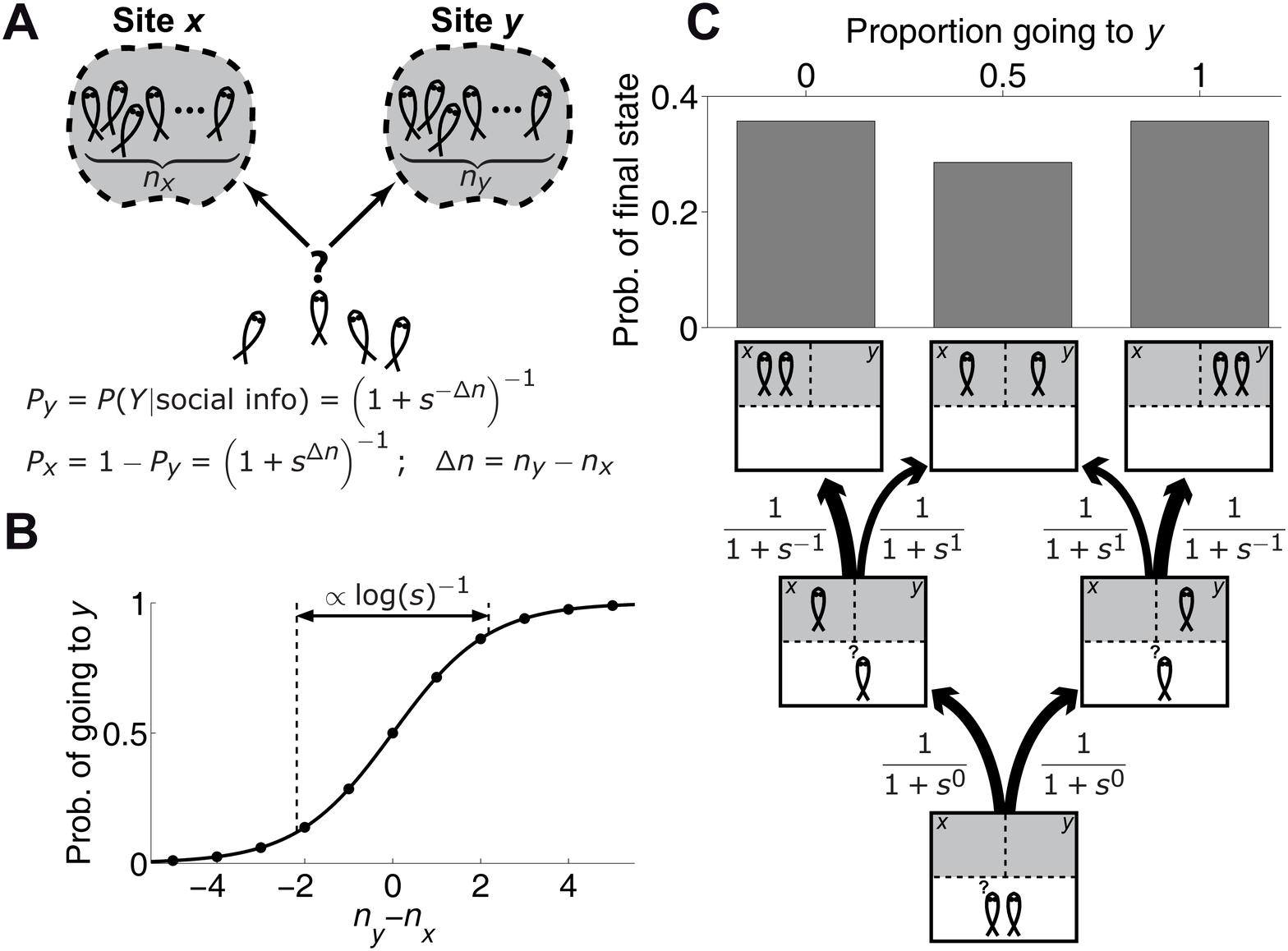}
\end{center}
\caption{
{\bf Model with individuals estimating which of two identical places is best.}  (\emph{A}) Schematic diagram of individuals choosing between two identical locations $\x$ and $\y$ when there are already $n_\x$ ($n_\y$) individuals at $\x$ ($\y$). (\emph{B}) Probability of going to $\y$ as a function of the difference between the number of individuals at $\y$ and $\x$, \refeq{3}. (\emph{C}) Sequential application of the behavioural rule in \refeq{3} with $\s=2.5$, for the simple case of a group of two individuals (bottom). The width of the arrows is proportional to the probability of each transition.  The 3 possible final configurations, with different proportion of individuals going to $\y$ (0, 0.5 and 1), have different probabilities of taking place, with both fish together at $\x$ or $\y$ being more probable than a group split (top).
}
\label{fig:1}
\end{figure}

We first considered the simple case of two identical equidistant sites, $\x$ and $\y$, \refig{1}{A}. For a set-up made symmetric by experimental design there is no true best option. But deciding individuals must act, like for any other case, using only their incomplete sensory data to make the best possible decision. Even when non-social sensory data indicates no relevant difference between the two sites, the social information can bias the estimation of the best option to one of the two sites. 

Using \refeq{general2choice} and that the three possible behaviors are `go to $\x$' ($\bx$), `go to $\y$' ($\by$) and `remain undecided' ($\beh{u}$), we obtain
\begin{equation}\label{ecn:general2choicesim}
P_y=\left(1+\ac\, \s_\x^{n_\x} \s_\y^{n_\y} \s_u^{N-n_\x-n_\y}\right)^{-1},
\end{equation}
where $n_\x$ and $n_\y$ are the number of individuals that have already chosen $\x$ and $\y$, respectively, and $N+1$ is the size of the group containing our focal individual and other $N$ animals. As the set-up is symmetric, the sensory information available to the deciding individual is the same for both options so $P(X|\ns)=P(Y|\ns)$ and then $\ac=1$ according to \refeq{a0}. Also, since indecision is not related to any particular choice, symmetry imposes $P(\beh{u}|X,\ns)=P(\beh{u}|Y,\ns)$, so indecision is not informative, $\s_u=1$ (\refeq{sk}). For the other two behaviors, going to $\x$ ($\bx$) and going to $\y$ ($\by$), \refeq{sk} gives
\begin{equation}
\begin{aligned}
&\s_\x=\frac{P(\bx|X,\ns)}{P(\bx|Y,\ns)}\\
&\s_\y=\frac{P(\by|X,\ns)}{P(\by|Y,\ns)}.
\end{aligned}
\end{equation}
$P(\bx|X,\ns)$ and $P(\by|Y,\ns)$ are the estimated probabilities of making the right choice, that is, going to $\x$ when $\x$ is the best option, or going to $\y$ when $\y$ is the best option. Since in this case the sensory information is identical for both options, the probability of making the correct choice must be the same for both options, $P(\bx|X,\ns)=P(\by|Y,\ns)$. An analogous argument holds for the incorrect choices, $P(\bx|Y,\ns)=P(\by|X,\ns)$, giving
\begin{equation}
\s_\x=1/\s_\y.
\end{equation}
In cases in which $\s_\x=1/\s_\y$, we find it convenient to express reliability more generally as
\begin{equation}
s\equiv \s_\x=1/\s_\y,
\end{equation}
which is the ratio of the probability of making the correct choice and the probability of making a mistake, for both behaviors. 
Using this definition and given that $\ac=a_u=1$,  \refeq{general2choicesim} reduces to
\begin{equation}\label{ecn:3}
P_\y=\left(1+\s^{-\Delta n}\right)^{-1},
\end{equation}
with the variable $\Delta n \equiv n_\y-n_\x$. \refeq{3} describes a sigmoidal function that is steeper the higher the higher the value of $\s$ (\refig{1}{B}). Therefore, for very reliable behaviors (high $\s$, meaning individuals that are much more likely to make correct choices than erroneous ones), $P_\y$ grows fast with $\Delta n$  and the deciding individual then goes to $\y$ with high probability when taking into account the behaviors of only very few individuals. 

The behavior of the group is obtained by applying the decision rule in \refeq{3} sequentially to each individual (see \meth). After each behavioural choice, we update the number of individuals at $\x$ and $\y$, using the new $n_\x$ and $n_\y$ for the next deciding individual (\refig{1}{C}, bottom). Repeating this procedure for all the individuals in the group, we can compute the probability for each possible final outcome of the experiment (\refig{1}{C}, top).

\begin{figure}[!ht]
\begin{center}
\includegraphics[width=\columnwidth]{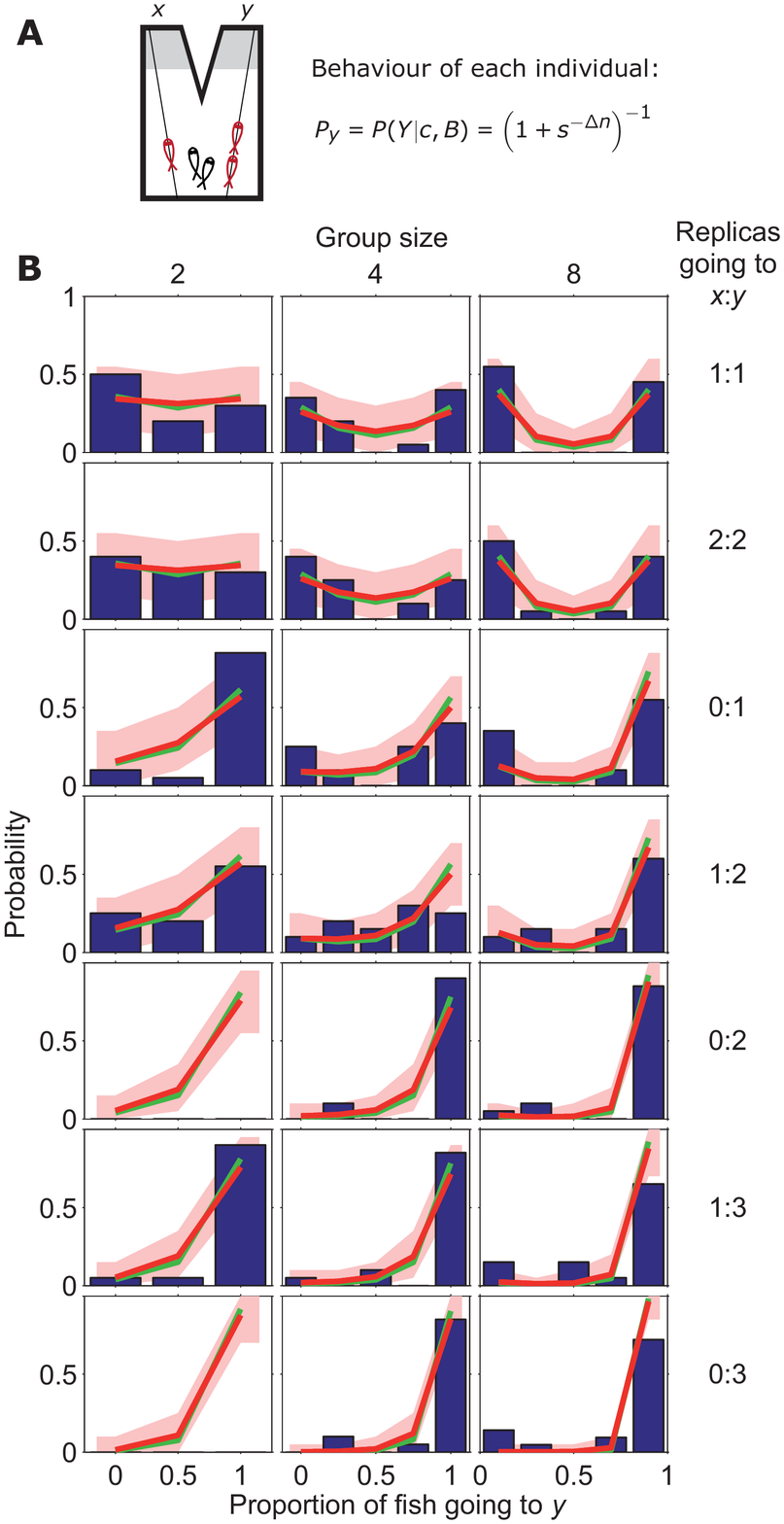}
\end{center}
\caption{
{\bf Comparison between model and stickleback choices in symmetric set-up.}  (\emph{A}) Schematic diagram of symmetric set-up with a group of sticklebacks (in black) choosing between two identical refugia and with different numbers of replica fish (in red) going to $\x$ and $\y$. (\emph{B}) Experimentally measured statistics of final configurations of fish choices from 20 experimental repetitions \cite{Ward2008} (blue histogram) and results from the model in \refeq{3} in the main text (red line using reliability parameter $\s=2.2$; red region: 95\% confidence interval; green line with $\s=2.5$). Different graphs correspond to different stickleback group sizes and different number of replicas going to $\x$ and $\y$.
}
\label{fig:2}
\end{figure}

\begin{figure}[!ht]
\begin{center}
\includegraphics[width=\columnwidth]{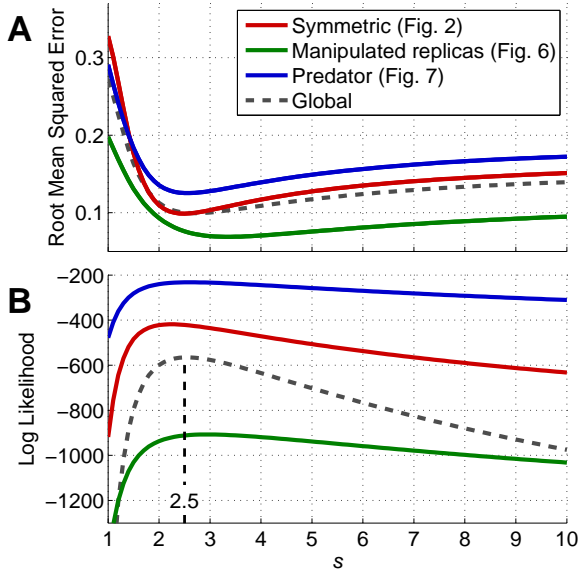}
\end{center}
\caption{
{\bf Goodness of fit for different values of the reliability ($\s$).}  \textbf{Red:} Symmetric case (plots in \refig{2}{}). \textbf{Green:} Case with different replicas at each side (plots in \refig{3}{}. The ratios $\s_\text{r}/\s_\text{R}$ are re-optimized for each value of $\s$). \textbf{Blue:} Asymmetric set-up with predator on one side (plots in \refig{4}{}; Parameter $\ac$ is re-optimized for each value of $s$). (\emph{A}) Root mean squared error between the data and the probabilities predicted by the model. Grey dashed line shows the mean RMSE for the three cases. The absolute values for each case depend on the shape of the data and are not comparable, only the trends and the position of the minima should be compared. (\emph{B}) Logarithm of the probability that the data come from the model. The height of each curve depends on the number of data for each experiment, only the trend and the position of the maxima should be compared. Grey dashed line shows the sum of the three coloured lines, but shifted by 1000 so that it fits on the scale. The peak of this global probability indicates the value of $\s$ that best fits the three datasets ($\s=2.5$).
}
\label{fig:S1}
\end{figure}

The relevance of the symmetric case is that the model has a single parameter and a single variable, enabling a powerful comparison against experimental data. We tested the model using an existing rich data set of collective decisions in three-spined sticklebacks \cite{Ward2008}, a shoaling fish species. This data set was obtained using a group of $\mo$ fish choosing between two identical refugia, one on their left and another one on their right (\refig{2}{A}), equivalent to locations $\x$ and $\y$ in the model (\refig{1}{A}). At the start of the experiment, $m_\x$  ($m_\y$) replica fish made of resin were moved along lines on the left (right) towards the refugia (\refig{2}{A}). The experimental results consisted on the statistics of collective decisions between the two refugia for 19 different cases using different group sizes $\mo$ = 2, 4 or 8 and different numbers of replicas going left and right, $m_\x:m_\y$ = \{1:1, 2:2, 0:1, 1:2, 0:2, 1:3, 0:3\} (\refig{2}{B}, blue histograms). To compare against these experimental data, we calculated the probability of finding a collective pattern applying the individual behavioural rule in \refeq{3} iteratively over each fish for the 19 experimental settings. We found a good fit of the model to the experimental data using for the 19 graphs the same value $\s=2.2$ (\refig{2}{B}, red line). The model is robust, with good fits in the interval $\s=\text{2-4}$ (\refig{S1}, red line). 

\begin{figure}[!ht]
\begin{center}
\includegraphics[width=\columnwidth]{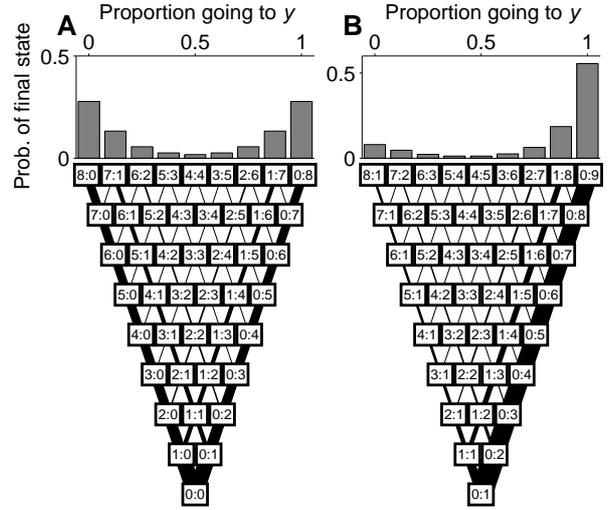}
\end{center}
\caption{
{\bf Illustration of the decision-making process in the model.}  \textbf{Bottom:} Decision-making process according to \refeq{3} (with $s=2.5$). Time runs from bottom to top. Each box represents a state with a given number of fish having already decided $\x$ or $\y$ ($n_\x:n_\y$). Each state can lead to another two states in the following time step, depending on whether the focal fish decides to go to $\x$ or $\y$. The width of the lines connecting states is proportional to the probability of that transition (equal to the probability of the prior state times the probability of the focal fish making the decision that leads to the later one). \textbf{Top:} Probability of each state after 8 fish have made their decisions. (\emph{A}) Case with no replicas, in which the final outcome is U-shaped. (\emph{B}) Case with one replica going to $\y$ (so initial state is already 0:1), in which the final outcome is J-shaped.
}
\label{fig:S2}
\end{figure}

Despite the simplicity of the behavioral rule in \refeq{3}, it reproduces the experimental results, including the dependence on the total number of fish $\mo$, even though the rule is independent of this parameter, except for determining the range of possible values of $\Delta n$. The dependence of the final distributions on $\mo$ emerges from the application of the rule to the $\mo$ individuals in the group, as is illustrated in \refig{S2}. Each small box represents a state of the system in which $n_x:n_y$ fish have already decided to go to $x$ and $y$, respectively. The lines connecting each box with another two boxes on top represent the decision made by the next deciding individual, that takes the system to the next state. The width of the lines is proportional to the probability of the decision. As more individuals decide, the central states become less likely simply because they accumulate more unlikely decisions. Therefore, the U-shape or J-shape becomes more pronounced for larger groups, even though the individual decision rule in \refeq{3} is independent of the total number of individuals $\mo$.

\begin{figure}[!ht]
\begin{center}
\includegraphics[width=\columnwidth]{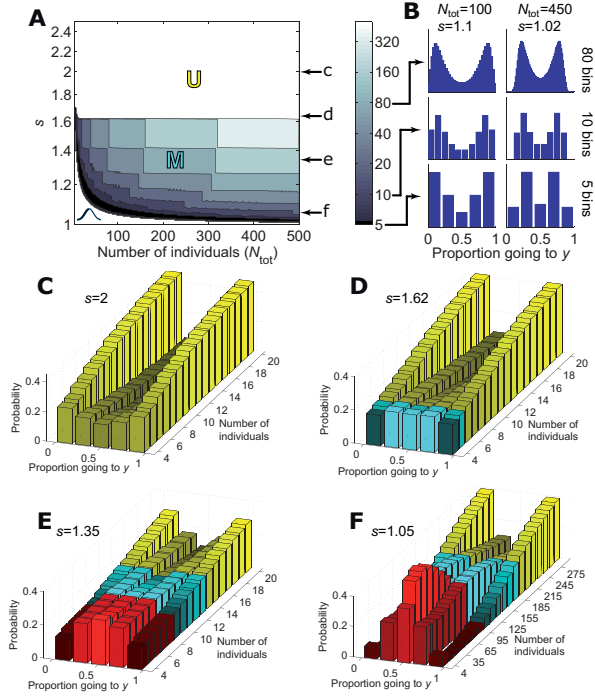}
\end{center}
\caption{
{\bf Types of distributions and dynamics for different values of reliability parameter $\s$ and group size.} (\emph{A}) Shape of histogram of final configurations as a function of $\s$ and the group size. Bell-shaped: white region on the bottom-left. M-shaped: region coloured in black and blue. As the observation of the M shape depends on the number of bins, the colorscale reflects the number of bins needed to observe the M shape (black has been reserved for exactly 5 bins). U-shape: white region on the top. Also, all the M-region above the black zone becomes U when the binning is too coarse. There is also a small region below the black zone where the M shape becomes a bell shape when the binning is too coarse. (\emph{B}) Dependence of the apparent shape on the number of bins: Top, 80 bins. Middle, 10 bins. Bottom, 5 bins. On the left, a probability that seems U-shaped for 5 bins, but is M shaped for a higher number of bins. On the right, a probability that stays M-shaped for any number of bins. (\emph{C-F}) Dynamics of the probability as the number of individuals increases for (\emph{C}) $\s=2$, (\emph{D}) $\s=1.62$, (\emph{E}) $\s=1.35$ and (\emph{F}) $\s=1.05$.
}
\label{fig:S3}
\end{figure}

Group decision-making in three-spined sticklebacks shows a single type of distribution in which probability is minimum at the center and increases monotonically towards the edges, denoted here as U-shaped distribution (or J-shaped when there is a bias to one of the two options). However, the model in \refeq{3} also gives two other types of distributions, \refig{S3}{A}. For non-social behavior ($\s=1$) the histogram is bell-shaped due to combinatorial effects. However, a bell-shape is also compatible with social animals for a certain range of $\s$ and group size (white region on the bottom-left of \refig{S3}{A}). For higher values of $\s$, the histograms are M-shaped, with two maxima located between the center and the sides (region coloured in black and blue in \refig{S3}{A}). However, the M shape becomes clear only with enough number of bins because the drop in probability near the edge or at the center of the distribution disappears when binning is too coarse, producing a bell-shaped or U-shaped histogram, \refig{S3}{B}. This is an important practical issue, because the amount of data that can be collected rarely allows for more than 5 bins. The colorscale in \refig{S3}{A} reflects the number of bins needed to observe the M shape (black has been reserved for exactly 5 bins). For high values of $\s$, the histograms are U-shaped (white region on the top of \refig{S3}{A}). Also, all the M-region above the black zone becomes of type U when the binning is too coarse. 

An interesting prediction of our model is that, for a given number of bins, the shape of the distribution of choices changes with the number of decided individuals, and the dynamics of this change depends on $s$. For high values of $s$, the probability is U-shaped from the beginning and becomes steeper as more individuals decide (as is the case for the stickleback dataset), \refig{S3}{C}. For lower values of $s$, we observe M-shaped distributions for the first individuals and then U-shaped ones when more individuals decide, \refig{S3}{D}. For even lower values of $s$, we observe bell-shaped distributions for the first individuals, then M-shaped and finally U-shaped, \refig{S3}{E,F}.

\subsection*{Symmetric set-up with modified replicas of animals}

\begin{figure}[!ht]
\begin{center}
\includegraphics[width=\columnwidth]{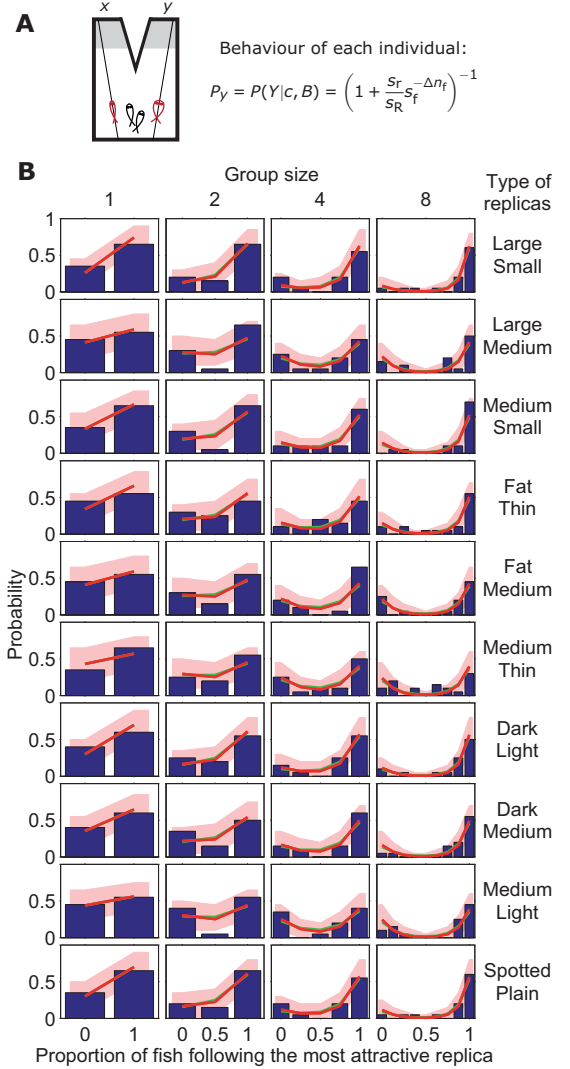}
\end{center}
\caption{
{\bf Comparison between model and stickleback choices with two differently modified replicas.}  (\emph{A}) Schematic diagram of symmetric set-up with a group of sticklebacks (in black) choosing between two identical refugia and with one replica fish going to $\x$ and a different one (in size, shape or pattern) going to $\y$ (in red). (\emph{B}) Experimentally measured statistics of final configurations of fish choices from 20 experimental repetitions \cite{Sumpter2008} (blue histogram) and results from model in \refeq{4} in the main text (red line using reliability parameter $\s_{\text{f}}=2.9$ and $\s_{\text{r}}/\s_{\text{R}}$ = 0.35, 0.7, 0.5, 0.52, 0.69, 0.75, 0.43, 0.55, 0.78, 0.43, for each row from top to bottom; red region: 95\% confidence interval; green line with $\s_{\text{f}}=2.5$ and same ratios $\s_{\text{r}}/\s_{\text{R}}$  as for red line). Different graphs correspond to different stickleback group sizes and different types of replicas going to $x$ and $y$.
}
\label{fig:3}
\end{figure}

An interesting modification of the experimental set-up consists in using replicas of the animals that we can modify to potentially alter their reliability estimated by the animals. We considered the particular case, motivated by experiments in \cite{Sumpter2008}, of two types of modified replicas with different characteristics (for example, fat or thin), \refig{3}{A}.  We considered 7 behaviors: `animal goes to $\x$' ($\beh{\text{f}\x}$), `animal goes to $\y$' ($\beh{\text{f}\y}$), `most attractive replica goes to $\x$' ($\beh{\text{R}\x}$), `most attractive replica goes to $\y$' ($\beh{\text{R}\y}$) `least attractive replica goes to $\x$' ($\beh{\text{r}\x}$), `least attractive replica goes to $\y$' ($\beh{\text{r}\y}$),  and 'animal remains undecided' ($\beh{\text{f}u}$). The probability of going to $\y$ in \refeq{general2choice} then reduces to
\begin{multline}\label{ecn:general2choice2replicas}
P_y=\\
\left(1+\ac\, \s_{\text{f}\x}^{n_{\text{f}\x}} \s_{\text{f}\y}^{n_{\text{f}\y}} \s_{\text{R}\x}^{n_{\text{R}\x}} \s_{\text{R}\y}^{n_{\text{R}\y}} \s_{\text{r}\x}^{n_{\text{r}\x}} \s_{\text{r}\y}^{n_{\text{r}\y}} \s_{\text{f}u}^{N_\text{f}-n_{\text{f}\x}-n_{\text{f}\y}}\right)^{-1},
\end{multline}
where subindex `f' refers to real fish and `R' (`r') to replicas of the most (least) attractive type. As in the previous section, symmetry imposes that $\ac=1$ and $\s_{\text{f}u}=1$. It also imposes the following relations between the reliability parameters, $\s_\text{f}\equiv\s_{\text{f}\x}=1/\s_{\text{f}\y}$, $\s_\text{R}\equiv\s_{\text{R}\x}=1/\s_{\text{R}\y}$, $\s_\text{r}\equiv\s_{\text{r}\x}=1/\s_{\text{r}\y}$. Therefore,
\begin{equation}
P_\y=\left(1+\s_{\text{f}}^{-\Delta n_\text{f}}\s_{\text{R}}^{-\Delta n_\text{R}}\s_{\text{r}}^{-\Delta n_\text{r}}\right)^{-1},
\end{equation}
where $\Delta n_\text{f} \equiv n_{\text{f}\y}-n_{\text{f}\x}$, $\Delta n_\text{R} \equiv n_{\text{R}\y}-n_{\text{R}\x}$ and $\Delta n_\text{r} \equiv n_{\text{r}\y}-n_{\text{r}\x}$.
In the particular case of only two different replicas, one going to $\x$ and the other to $\y$ and for notational simplicity taking the convention that the most (least) attractive replica goes to $\y$ ($\x$), we have $\Delta n_\text{R}=1$ and $\Delta n_\text{r}=-1$. Therefore,
\begin{equation}\label{ecn:4}
P_\y=\left(1+\frac{\s_{\text{r}}}{\s_{\text{R}}}\s_{\text{f}}^{-\Delta n_\text{f}}\right)^{-1}.
\end{equation}
Note that the probability in \refeq{4} does not depend on $\s_{\text{r}}$ and $\s_{\text{R}}$ separately, but only on their ratio. Therefore, in this case the model uses only two parameters ($\s_{\text{f}}$ and $\s_{\text{r}}/\s_{\text{R}}$). We compared the model with the stickleback data set from \cite{Sumpter2008}, \refig{3}. The data in \refig{3}{B} has a different type of replica pair in each row, so in principle we would fit a different ratio $\s_{\text{r}}/\s_{\text{R}}$ for each row. But note that the first three rows correspond to experiments with the same three replicas (large, medium and small), combined in different pairs. The same can be said for the second and third threesomes of rows. Therefore, there are only two free parameters for each three rows. On the other hand, $\s_{\text{f}}$ should have the same value for all cases. The model again reproduces the experimental results reported in reference \cite{Sumpter2008} , obtaining the best fit for $\s_{\text{f}}=2.9$ (\refig{3}{B}). The result is robust, with good fits for $\s_{\text{f}}=\text{2-4}$ (\refig{S1}, green line) in accord with the value obtained for the case shown in \refig{2}{B}. 

\subsection*{Asymmetric set-up}
We finally considered the case in which sites $\x$ and $\y$ are different and the three behaviors are `go to $\x$' ($\beh{\x}$), `go to $\y$' ($\beh{\y}$) and `remain undecided' ($\beh{u}$). \refeq{general2choice} reduces to 
\begin{equation}\label{ecn:general2choiceasim}
P_y=\left(1+\ac\, \s_\x^{n_\x} \s_\y^{n_\y} \s_u^{N-n_\x-n_\y}\right)^{-1}.
\end{equation}
The term $\ac=P(\X|\ns)/P(\Y|\ns)$ represents the non-social information and in general $\ac\neq 1$ because the set-up is asymmetric by design. This asymmetry might also affect how a deciding animal takes into account the behaviours of other animals depending on which side they chose, making in general $\s_\x\neq 1/\s_\y$. Also, indecision might be informative. For example, if non-social information indicates the possible presence of a predator at $\y$, the indecision of other animals might confirm this to the deciding individual, further biasing the decision towards $\x$. Therefore, we may have $\s_u\neq 1$. But it may also be the case that the set-up's asymmetry does not affect the social terms, so we also tested a simpler model in which $s\equiv\s_x=1/\s_y$ and $\s_u=1$, giving 
\begin{equation}\label{ecn:5}
P_\y=\left(1+\ac\, \s^{-\Delta n}\right)^{-1}.
\end{equation}

\begin{figure}[!ht]
\begin{center}
\includegraphics[width=\columnwidth]{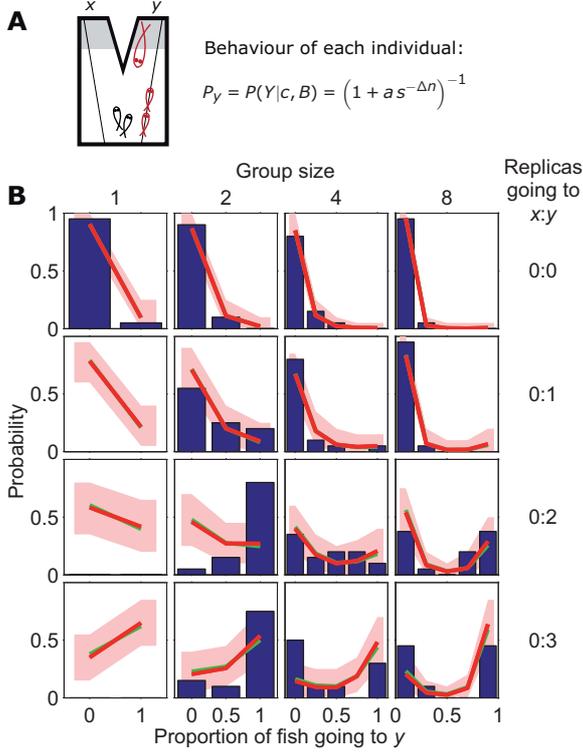}
\end{center}
\caption{
{\bf Comparison between model and stickleback choices in asymmetric set-up.}  (\emph{A}) Schematic diagram of asymmetric set-up (predator at $\y$, large fish depicted in red) with a group of sticklebacks (in black) choosing between two refugia, and replica fish (small fish depicted in red) going to $\y$. (\emph{B}) Experimentally measured statistics of final configurations of fish choices from 20 experimental repetitions \cite{Ward2008} (blue histogram) and results from model in \refeq{5} in the main text (red line using $\s=2.6$, $\ac=9.5$; red region: $95\%$ confidence interval. Green line using $\s=2.5$ and same $\ac$ as for red line). Different graphs correspond to different stickleback group sizes and different number of replicas going to $\y$.
}
\label{fig:4}
\end{figure}

The stickleback dataset reported in reference \cite{Ward2008} is ideally suited to test the asymmetric model for the experiments that were performed with a replica predator at the right arm (\refig{4}{A}). The model in \refeq{5} fits best the data with $\s=2.6$ (\refig{4}{B}) and it is robust with a good fit in $\s=\text{2-4}$ (\refig{S1}, blue line). The more complex model in \refeq{general2choiceasim} gives fits very similar to those of simpler model. Specifically, parameter $\s_u$ was rejected by the Bayes Information Criterion \cite{Schwarz1978,Link2006}, suggesting that fish do not rely on undecided individuals. The fact that fish rely differently on other fish depending on the option they have taken could not be ruled out by the Bayes Information Criterion, but in any case the impact of this difference on the data is small.

In the experiments in \refig{2}{} and \refig{4}{}, we have assumed that the replicas are perceived by fish as real animals. However, it is reasonable to think that fish might perceive the difference, and rely differently on replicas and real fish. To test this, we considered different behaviors for fish and replicas, such as `fish goes to $\x$' and `replica goes to $\x$'. Making that distinction, we get that \refeq{general2choice} reduces to
\begin{equation}\label{ecn:supergeneral2choiceasim}
P_y=\left(1+\ac\, \s_{\text{f}\x}^{n_{\text{f}\x}} \s_{\text{f}\y}^{n_{\text{f}\y}} \s_{\text{r}\x}^{n_{\text{r}\x}} \s_{\text{r}\y}^{n_{\text{r}\y}} \s_{\text{f}u}^{N_\text{f}-n_{\text{f}\x}-n_{\text{f}\y}}\right)^{-1}.
\end{equation}
The Bayes Information Criterion rejects only parameter $\s_{\text{f}u}$. However, the addition of the new parameters that distinguish replica from real fish give very small improvements in the fits compared to results of the simpler models in \refeq{3} and \refeq{5} (see \refig{S4}{} and \ref{fig:S6}), suggesting that fish follow replicas as much as they follow real fish. 

\subsection*{Model including dependencies}
\begin{figure}[!ht]
\begin{center}
\includegraphics[width=\columnwidth]{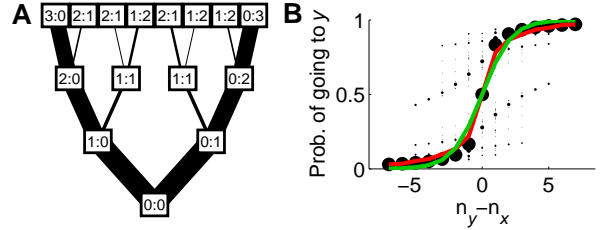}
\end{center}
\caption{
{\bf Model taking into account dependencies.} (\emph{A}) Decision-making process according to the model with dependencies, \refeq{productorio}-\textbf{\ref{ecn:SKfinal}}. Time runs from bottom to top. Each box represents one state, and each edge represents one option of the deciding individual, that either goes to $x$ or to $y$. Edge width is proportional to the probability of the decision. (\emph{B}) Probability of choosing $y$ as a function of the difference of the number of individuals that have already chosen each option ($\Delta n=n_\y-n_\x$), for $a_\X'=5$. In the new model the probability does not depend any more on $\Delta n$ alone, so states with the same $\Delta n$ have different values for the probability (black dots). The area of the dots is proportional to the probability of observing each state. Red line shows the expected value of the probability for each value of $\Delta n$. The green line shows the probability for the model that neglects dependencies (\refeq{3}), $\left(1+s^{-\Delta n}\right)^{-1}$ for $\s=2.5$.
}
\label{fig:modelocaminos}
\end{figure}

In this section we will remove the hypothesis of independence among the behaviors of the other individuals (\refeq{behavindep}). We now consider that the focal individual not only takes into account the behaviors of the other animals at the time of decision but the specific sequence of decisions that has taken place before, $\{b_i\}_{i=1}^{K-1}$, being $K-1$ the number of individuals that have decided before the focal one. For example, the sequence $\{\x,\y\}$ may give different information to the focal individual than the sequence $\{\y,\x\}$. This is illustrated in \refig{modelocaminos}{A}, where there are two possible paths leading to states labeled as 1:1, but these two states are in different branches of the tree (in contrast with \refig{S2}{}, in which these two states were collapsed in a single one).

To calculate the probability of the observed sequence of behaviors provided that $\Y$ is the correct choice, 
\begin{equation}
P(\{b_i\}_{i=1}^{K-1}|\Y,\ns_K),
\end{equation}
one can apply $P(A,B)=P(A|B)P(B)$ repeatedly to obtain
\begin{equation}\label{ecn:productorio}
P(\{b_i\}_{i=1}^{K-1}|\Y,\ns_K)=\prod_{k=1}^{K-1}P(b_k|\Y,\ns_K,\{b_i\}_{i=1}^{k-1}).
\end{equation}
This expression substitutes the assumption of independence in \refeq{behavindep}. Each of the terms in the product is simply the probability that the $k\text{-th}$ individual makes its decision, given the previous decisions, and also given that $y$ is the correct choice. This result was expected since if we look at the tree in \refig{modelocaminos}{A} we see that the probability of reaching a given state is simply the product of the probabilities of choosing the adequate branches in each step.

So the problem reduces to computing the individual decision probabilities $P(b_k|\Y,\ns_K,\{b_i\}_{i=1}^{k-1})$. We assume in the following that these probabilities are calculated by the focal individual by assuming that all animals use the same rules to make a decision. The rule for the focal individual is, as in previous sections,
\begin{equation}\label{ecn:PYK}
P_{y_K}=P(Y|\{b_i\}_{i=1}^{K-1},\ns_K)=\frac{1}{1+\ac_K\, S_K},
\end{equation}
where the non-social and social terms are
\begin{equation}\label{ecn:a0K}
\ac_K=\frac{P(\X|\ns_K)}{P(\Y|\ns_K)},
\end{equation}
and
\begin{equation}\label{ecn:SK}
S_K=\frac{P(\{b_i\}_{i=1}^{K-1}|\X,\ns_K)}{P(\{b_i\}_{i=1}^{K-1}|\Y,\ns_K)},
\end{equation}
respectively, and where we have added subscript $K$ to $S$, $\ac$ and $\ns$ to reflect that they apply to the focal individual, that makes its decision in the $K\text{-th}$ place. 

The assumption that all animals apply the same rules translates into the following. To apply an equation like \refeq{PYK} but on a different individual (say, individual $k$) it is necessary to know the non-social information $\ns_k$. Remember that all these computations are made from the point of view of the focal individual, and obviously the focal individual does not have access to the non-social information of the other individuals. It may seem reasonable for the focal animal to assume that all the other individuals have the same non-social information ($\ns_K$), but this would result in no social behavior at all (if the other individuals have the same non-social information, their behaviors will not give any extra information). Instead, one can assume that the other individuals may have a different non-social information, $\ns'$. Furthermore, this non-social information depends on which is the best choice, because if for example $\x$ is the best choice the other individuals have some probability of detecting it, and therefore their non-social information will be on average biased towards $\x$. We approximate this average bias by assuming that, if $\x$ ($\y$) is the best choice, all the other individuals will have non-social information $\ns_\X'$ ($\ns_\Y'$) that will bias the decision towards $\x$ ($\y$). It is therefore the same to assume that $\x$ ($\y$) is the best option as to assume that all the other individuals have non-social information $\ns_\X'$ ($\ns_\Y'$). Therefore, for the probabilities of individual behaviors in \refeq{productorio}, we have that
\begin{equation}
P(b_k|\Y,\ns_K,\{b_i\}_{i=1}^{k-1})=P(b_k|\ns_\Y',\ns_K,\{b_i\}_{i=1}^{k-1}),
\end{equation}
where now $c_Y'$ applies to the $k\text{-th}$ individual, so we can compute this probability simply by applying \refeq{PYK} to the $k\text{-th}$ individual,\begin{equation}\label{ecn:PYKX}
P_{y_k,Y}=\frac{1}{1+\ac_Y'\, S_k},
\end{equation}
where
\begin{equation}\label{ecn:a0X}
a_\Y'=\frac{P(\X|\ns_\Y')}{P(\Y|\ns_\Y')}.
\end{equation}
Then, if we denote $P_{b_k,Y}\equiv P(b_k|\ns_\Y',\ns_K,\{b_i\}_{i=1}^{k-1})$, we have that
\begin{equation}
\left\{
\begin{array}{ll}
P_{b_k,Y}=P_{y_k,\Y} &\text{ if } b_k=y\\
P_{b_k,Y}=1-P_{y_k,\Y} &\text{ if } b_k=x.
\end{array}
\right.
\end{equation} 
These are the individual probabilities needed in \refeq{productorio}, that takes into account the correlations among the other individuals. So we can already calculate $S_k$ using \refeq{SK}, 
\begin{equation}\label{ecn:SKfinal}
S_k=\frac{\prod_{i=1}^{k-1}P_{b_i,X}}{\prod_{i=1}^{k-1}P_{b_i,Y}},
\end{equation}
Eqs. \textbf{\ref{ecn:PYKX}} and \textbf{\ref{ecn:SKfinal}} have a recursive relation, because we need the probabilities up to step $k-1$ to compute $S_k$, and then we need $S_k$ to compute the probabilities in step $k$. At the beginning no individual has made any choices, so we start with $S_1=1$ and work recursively from there until we obtain the probabilities for individual $K-1$, that allow to compute $S_K$. Then, we can already use \refeq{PYK} to compute the decision probability of the focal individual, this time using its actual non-social term $a_K$ (which is 1 for the symmetric cases, and fitted to the data in the non-symmetric case).

The equations above constitute the model taking into account dependencies. The new parameters of this model are $\ac_\X'$ and $\ac_\Y'$, which substitute $s_\x$ and $s_\y$ in the previous models, so the number of parameters is exactly the same. In the symmetrical case we must have that $\ac_\X'=1/\ac_\Y'$, so the model has a single parameter. For the non-symmetrical case these parameters may be independent of each other, but we find good results even assuming that they are not, as was the case for the simplified model. So for simplicity we always assume that
\begin{equation}
\ac_\X'=1/\ac_\Y'.
\end{equation}
For the case with different replicas at each side, each of them has a different value of $a_\X'$, thus making one replica more attractive than the other.

\begin{figure}[!ht]
\begin{center}
\includegraphics[width=.8\columnwidth]{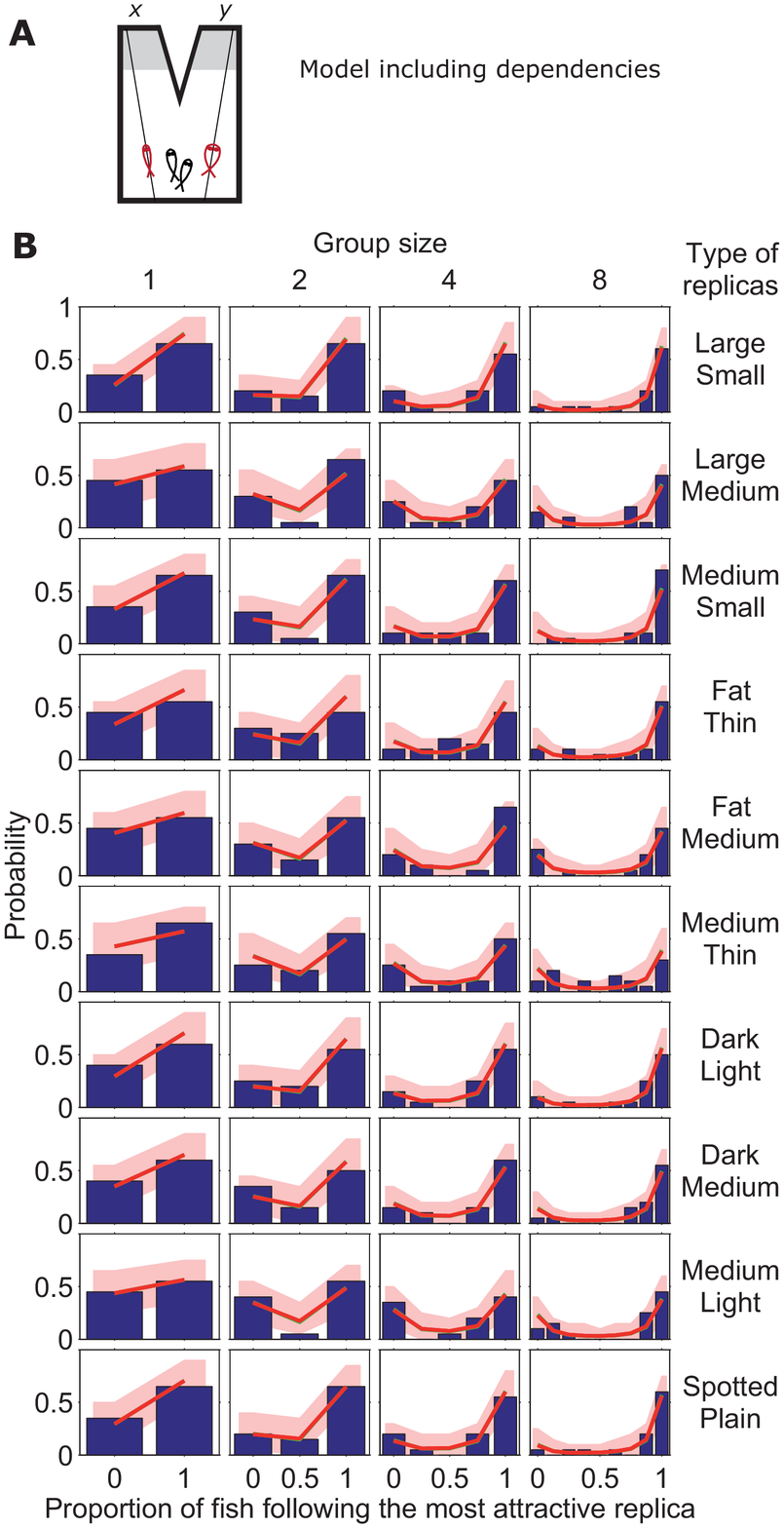}
\end{center}
\caption{
{\bf Comparison between model including dependencies and stickleback choices with two differently modified replicas.}  (\emph{A}) Schematic diagram of symmetric set-up with a group of sticklebacks (in black) choosing between two identical refugia and with one replica fish going to $\x$ and a different one (in size, shape or pattern) going to $\y$ (in red). (\emph{B}) Experimentally measured statistics of final configurations of fish choices from 20 experimental repetitions \cite{Sumpter2008} (blue histogram) and results from model that takes dependencies into account (red line, with $\ac_{\X,\text{fish}}'=4.8$ and $\ac_{\X,\text{replicas}}'$ = 21.4, 11.8, 0.6, 9.9, 4.8, 0.9, 13, 8, 0.7, 14.5, 0.9, for each type of replica (large, medium, small, fat, etc.); red region: 95\% confidence interval; green line with $\ac_{\X,\text{fish}}'=5$ and same $\ac_{\X,\text{replicas}}'$ as for red line). Different graphs correspond to different stickleback group sizes and different types of replicas going to $x$ and $y$.
}
\label{fig:datosCurrBiolcaminos}
\end{figure}

The new model also matches very well with the experimental data discussed in this paper. Results for the case of two different replicas are shown in \refig{datosCurrBiolcaminos}{}, for the symmetric case in \refig{datoswardnopredcaminos}{} and for the case with predator in \refig{datoswardpredcaminos}{}. Fits are robust, and all cases are well explained by the model with the same value of $a_\X'=5$, \refig{fita1_caminos}. See Figs. \ref{fig:S4}{}-\ref{fig:S6} for a comparison of all models.

We now ask how different is the model including dependencies from the model that neglects them. To compare the two models, we plot the probability of going to $\y$ as a function of $\Delta n=n_\y-n_\x$ for the new model, as we did in \refig{1}{B} for the old one. The inclusion of dependencies has the consequence that the probability of going to $\y$ does not depend only on $\Delta n$, since now different states with the same $\Delta n$ may have different probabilities. Therefore, when we plot the probability of going to $\y$ as a function of $\Delta n$ we obtain different values of the probability for each value of $\Delta n$. This is shown by the black dots in \refig{modelocaminos}{B}, where the size of the dots is proportional to the probability of observing each state when starting from 0:0. The red line shows the average probability for each $\Delta n$, taking into account the probability of each state. Both the dots and this line correspond to $\ac_\X'=5$, which is the one that fits best the data. The green line corresponds to the probability for the simplest model neglecting dependencies, with the value that best fits to the data ($\s=2.5$). This line is close to the mean probability for the new model and to the values with highest probability of occurrence, so the simple model is as a good approximation to the model with dependencies.

We find an interesting prediction of the new model: There are some states in which the most likely option is to choose the option chosen by \emph{fewer} individuals (for example, note in \refig{modelocaminos}{D} that some points with $\Delta n<0$ are above 0.5). This surprising result comes from the fact that, as more fish accumulate at one side, their choices become less and less informative (because it is very likely that they are simply following the others). If then one fish goes to the opposite side, its behavior is very informative, because it is contradicting its social information. This effect can be so strong that it may beat the effect of all the other individuals, resulting in a higher probability of following this last individual than all the individuals that decided before. 

\section*{Discussion}

We have shown that probabilistic estimation in the presence of uncertainty can explain collective animal decisions. This approach generated a new expression for each experimental manipulation, Eq. \textbf{\ref{ecn:3}}-\textbf{\ref{ecn:5}}, and was naturally extended to test for more refined cognitive capacities, \refeq{supergeneral2choiceasim}. The model was found to have a good correspondence with the data in three experimental settings (Figs. \ref{fig:2}, \ref{fig:3} and \ref{fig:4}), always giving a good fit with the social reliability parameter $\s$ in the interval 2-4. Indeed, all the data have a very good fit with $\s=2.5$ (Figs. \ref{fig:2}, \ref{fig:3} and \ref{fig:4}, green lines). According to \refeq{sk}, this value for $\s$ has the interpretation that, for the behaviors relevant for these experiments, the fish assume that their conspecifics make the right choice 2.5 times more often than the wrong choice. 

For the data used in this paper, previous empirical fits used more parameters \cite{Ward2008} (Figs. \ref{fig:S4}-\ref{fig:S6}, blue
line), and added more complex behavioral rules when the basic model failed \cite{Sumpter2008} (\refig{S5}, blue line). Our approach thus gains in simplicity. It also finds an expression for each set-up with expressions for complex set-ups obtained with add-ons to those of simpler set-ups, making the model scalable and easier to understand in terms of simpler experiments. Also, taking the models as fits to experimental data, the bayesian information criterion finds our models to be better than those in \cite{Ward2008} and \cite{Sumpter2008} (see captions in Figs. \ref{fig:S4}-\ref{fig:S6} for details).   

Collective animal behavior has been subject to a particularly careful quantitative analysis. Previous studies have given descriptions led by the powerful idea that complex collective behaviors can emerge from simple individual rules. In fact, some systems have been found empirically to obey rules that are mathematically similar or the same as some of the ones presented in this paper, further supporting the idea that probabilistic estimation might underlie collective decision rules in many species. For example, a function like the one in \refeq{3} has been used to describe the behavior of Pharaoh's ant \cite{Jeanson2003}, a function like \refeq{5} for mosquito fish \cite{Ward2011}, and a function like the one in the right-hand-side of \refeq{5} for meerkats \cite{Bousquet2010}. But despite the importance of group decisions in animals, little is known about the origin of such simple individual rules. This paper argues that probabilistic estimation can be an underlying substrate for the rules explaining collective decisions, thus helping in their evolutionary explanation. Also, this connection between patterns in animal collectives and a cognitive process helps to explain the similarities that exist between decision-making processes at the level of the brain and at the level of animal collectives \cite{Marshall2009,Couzin2009}.

Our model is naturally compatible with other theories that use a Bayesian formalism to study different aspects of behavior and neurobiology, thus contributing to a unified approach of information processing in animals. For example, it may be combined with the formalism of Bayesian foraging theory \cite{OATEN1977}, through an expansion of the non-social reliability $\ac$. Related to this case, a very well studied example of use of social information is the one in which one individual can observe directly the food collected by another individual \cite{Valone1989,Templeton1995,Templeton1996,Smith1999,CLARK1986}. In this case the social information is as unambiguous as the non-social one, so in this case both types of information should have a similar mathematical form \cite{Valone1989,Templeton1995,Templeton1996,Smith1999,CLARK1986}. This is consistent with our model, that in this case will give a similar expression for $a$ and $S$. Other kinds of social information (such as another individual's decision to leave a food patch or choices of females in mating \cite{Nordell1998}) would enter naturally in our reliability terms $\s_k$. In discussing these and similar problems, it has been proposed that animals should use social information when their personal information is poor, and ignore it otherwise \cite{Dall2005,Giraldeau2002,Nordell1998}. Our model provides a quantitative framework for this problem, predicting that social information is always used, only with different weights with respect to other sources of information. Bayesian estimation is also a prominent approach to study decisions in neurobiology and psychology \cite{Helmholtz1925,Mach1980,Knill2004,Jacobs1999,Knill2003a,Ernst2002,Battaglia2003,Alais2004,Gold2001,Kording2004,Kording2006,Gold2007,Courville2006,Kruschke2006,Tenenbaum2011} and it would be of interest to explore the mechanisms and role played by the multiplicative relation between non-social and social terms. 

Our approach also makes a number of predictions. For example, it derives the probability of choosing among $M$ options (see Eq. \ecnSXVI\ of the \supptext), that for the symmetric case reduces to
\begin{equation}\label{ecn:generalMchoices}
P_{\mu}=\left(1+\sum_{m \neq \mu}^{M} \s^{-\left(n_{\mu}-n_{m}\right)}\right)^{-1},
\end{equation}
predicted also to fit the data for cases with $M>2$ options.

We also predict a quantitative link between estimation and collective behavior. The parameters $\ac$ and $\s_k$ in our model are in fact not merely fitting parameters, but true experimental variables. Manipulations of $\ac$ and $\s_k$ should allow to test that changes in collective behavior follow the predictions of the model. A counterintuitive prediction about the manipulation of $\s_k$ is that external factors unrelated to the social component can nevertheless modify it. For example, a fish that usually finds food in a given environment should interpret a sudden turn of one of his mates as an indication that it has found food, and therefore will follow it. In contrast, another fish that is not expected to find food in that environment will not interpret the sudden turn as indicative of food, and will not follow. Thus, the model predicts that the \emph{a priori} probability of finding food (to which each fish can be trained in isolation) will modify its propensity to follow conspecifics. An alternative approach that would not need manipulation of the reliabilities $\s_k$ would consist in showing that the probability of copying a behavior increases with how reliably the behavior informs about the environment.

We can also extend the estimation model to use, instead of the location of animals, their predicted location. We would then find expressions like the ones in this paper but for the number or density of individuals estimated for a later time. Consider for example the case without non-social information, described in \refeq{3} for two options and in \refeq{generalMchoices} for more options. We can rewrite these equations as $P_\tref=\Omega \s^{n_\tref}$ with $\tref$ one of the options and $\Omega$ is the normalization, $\Omega=\sum_{m=1}^M s^{n_m}$, where $M$ is the number of options. Then, we would have $P(\vec{x}) =\Omega s^{\rho(\vec{x}; t + \Delta t)}$ for the continuous case using prediction. Future positions at times $t + \Delta t$ (where $\Delta t$ does not need to be constant) in terms of variables at present time $t$ would be given by $\vec{x} + \vec{v} \Delta t$ for animals moving at constant velocity $\vec{v}$. Consider then a simple case of an animal located at $\vec{x}$ and estimating the future position of a compact group at $\vec{x}_g$ and moving with velocity $\vec{v}_g$. The deciding animal would be predicted to move with a high probability in the direction $\left(\vec{x}_{g}(t)-\vec{x}(t)\right) + \Delta t \vec{v}_{g}(t)$. Estimation of future locations thus naturally predicts in this simple case a particular form of `attraction' and `alignment' forces of dynamical empirical models \cite{Couzin2005,Couzin2010} as  attraction to future positions, but in the general also deviations from these simple rules.



\section*{Methods}

\subsection*{Obtaining group behavior from the model of an individual}
The estimation rules presented in this paper refer to a single individual. To simulate the behavior of a group, we use the following algorithm: The current individual decides between $\x$ and $\y$. After the decision, we recompute the relevant parameters of the model and use the new values for the next deciding individual. The undecided individuals are only those that are waiting for their turn to decide. We tested an alternative implementation in which individuals may remain undecided or in which two individuals can decide simultaneously, obtaining no relevant differences. 

For the case of the model including dependencies, the model always starts at state 0:0, with $S=1$. Most experiments have initial conditions in which several replicas are already going to either side, and the fish have no information about the path followed to reach this state. In these cases, we average the probabilities of all the paths that might have possibly led to the initial state to compute the initial value of $S$.

\suppdata, contain Matlab functions that run the models (extensions of the files must be changed from .txt to .m to make them operative). \emph{Protocol S1} corresponds to the model without dependencies, and \emph{Protocol S2} corresponds to the model with dependencies. These functions have been used to generate all the theoretical results presented in this paper.

\subsection*{Fits}
We computed log likelihood as the logarithm of the probability that the histograms come from the model. We searched for the model parameters giving a higher value of log likelihood, corresponding to a better fit. This search was performed by optimizing each parameter separately (keeping the rest constant)  and iterating through all parameters until convergence. In all cases convergence was rapidly achieved. We performed multiple searches for best fitting parameters starting from random initial conditions and always found convergence to the same values, suggesting there are no local maxima. Indeed, we observed that log-likelihood is smooth and with a single maximum in all the cases with 1 or 2 parameters (see \refig{S1}{} for an example).

\subsection*{Bayesian Information Criterion}

For model comparison we used the Bayesian Information Criterion (BIC) \cite{Schwarz1978,Link2006}, which takes into account both goodness of fit and the number of parameters. According to this criterion, among several models that have been fitted to maximize log likelihood, one should select the one for which 
\begin{equation}\label{ecn:15}
BIC_i=L_i-\frac{1}{2}k_i\log(h)
\end{equation}
is largest, where $L_i$ is the logarithm of the probability that the data comes from the $i\text{-th}$ model once its parameters have been optimized to maximize this probability, $k_i$ is its number of parameters of the $i\text{-th}$ model and $h$ is the number of measurements (which in our case is the same for all models).

More intuitive than the direct $BIC_i$ values in \refeq{15} are the BIC weights, defined as \cite{Link2006}
\begin{equation}\label{ecn:BICweigths}
w_i=\frac{\exp(BIC_i)}{\sum_j{\exp(BIC_j)}},
\end{equation}
when we assume that all models are \emph{a priori} equally likely. Roughly speaking, $w_i$ can be interpreted as the probability that model $i$ is the most correct one \cite{Link2006}.

We used BIC to compare different versions of our model, and also to compare our model with those of references \cite{Ward2008,Sumpter2008} (see Figs. \ref{fig:S4}-\ref{fig:S6}). The models of refs. \cite{Ward2008,Sumpter2008} were originally fitted by minimizing the mean squared error instead of by maximizing logprob. For this reason, they score very poorly in BIC with their reported parameters. For this reason, we re-optimized for maximum logprob all their model parameters (these parameters are, using the notation of refs. \cite{Ward2008,Sumpter2008}, $a$, $k$, $T$, $m$ and $r$, with $r$ only applicable in the case of predator present). For the case of different replicas going to each side, parameter $p_\text{bias}$ takes a different value for each row in the figure, adding up to 10 parameters. The model in ref. \cite{Sumpter2008} is computationally expensive, so it is not feasible to re-optimize these many parameters. Therefore, we treated them as if they were independently measured: we fixed $p_\text{bias}$ in each case so that the results of the trials with a single individual matched exactly the model's prediction (as reported in \cite{Sumpter2008}). We also followed this procedure with the ratios $\s_{\text{r}}/\s_{\text{R}}$ of our model without dependencies, and the pairs $\ac_{\X,\text{replicas}}'$ in our model with dependencies. Then, we performed BIC taking into account neither these parameters ($p_\text{bias}$ the ratios $\s_{\text{r}}/\s_{\text{R}}$ and the pairs $\ac_{\X,\text{replicas}}'$) nor the data from trials using single individuals.

\section*{Acknowledgments}
We acknowledge useful comments by Sara Arganda, Larissa Conradt, Iain Couzin, Jacques Gautrais, David Sumpter, Guy Theraulaz, Juli\'an Vicente Page and COLMOT 2010 participants.\\ 

\onecolumn
\bibliography{Referencias}{}

\clearpage
\section*{Supporting Figures} 
\renewcommand{\thefigure}{S\arabic{figure}}
\setcounter{figure}{0}

\begin{figure}[!ht]
\begin{center}
\includegraphics[width=0.4\columnwidth]{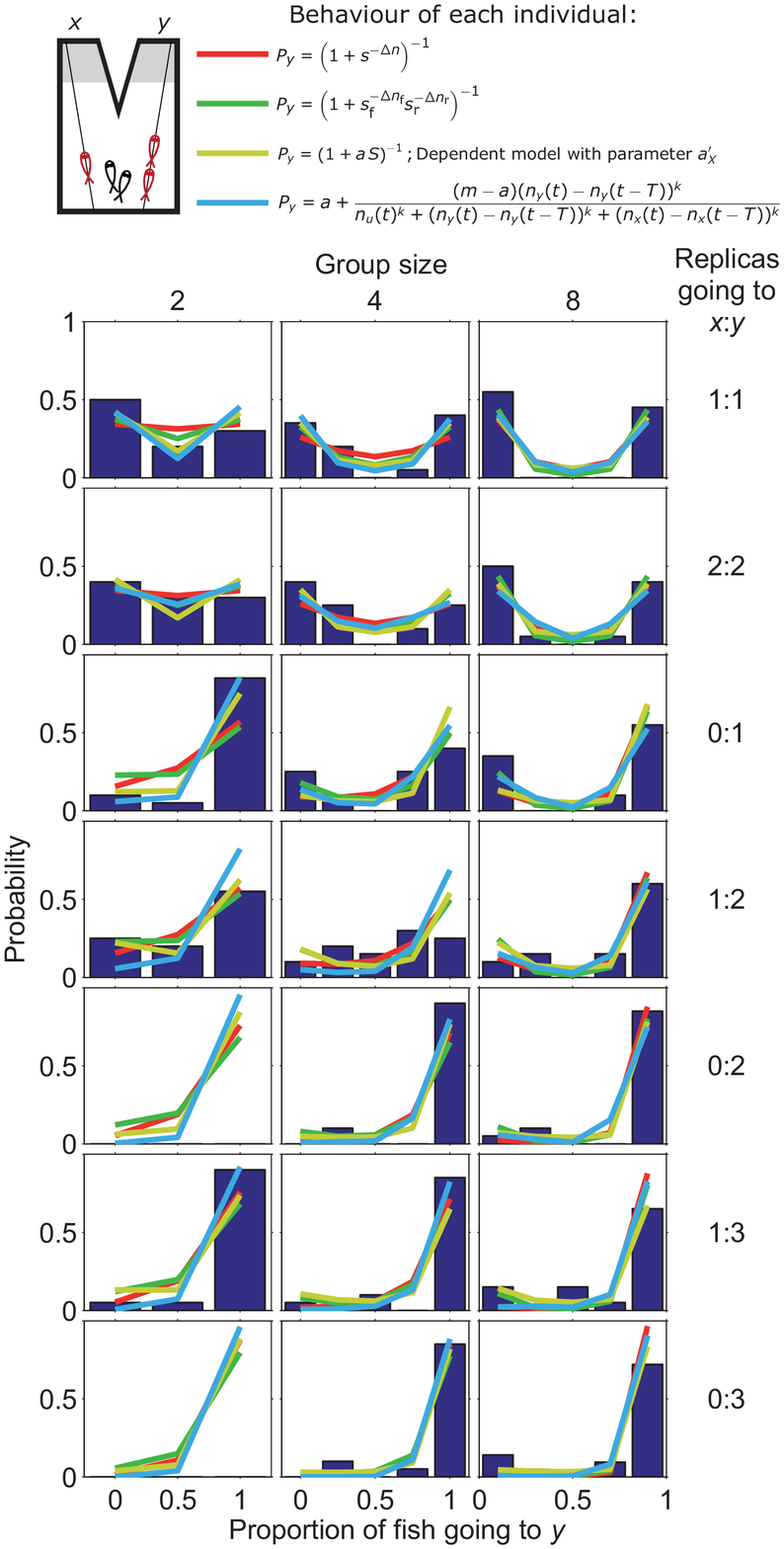}
\end{center}
\caption{
{\bf Comparison between different models for the symmetric set-up.} Experimentally measured statistics of final configurations of fish choices from 20 experimental repetitions \cite{Ward2008} (blue histograms). Red line: results from our single-parameter model assuming independence in \refeq{3} in the main text ($\s=2.2$\comment{; red region: 95\% confidence interval}). Green line: Enhanced model assuming independence with different reliability for the replicas ($\s_\text{f}=3$, $\s_\text{r}=1.76$). Yellow line: Model including dependencies ($\ac_\X'=4.9$). Blue line: Empirical model presented in Ref. \cite{Ward2008}, using the parameters reported there. Different graphs correspond to different stickleback group sizes and different number of replicas going to $\x$ and $\y$. According to Bayesian Information Criterion (BIC, see \meth), the best model is our model with dependencies (yellow line, logprob $L=-394$, and BIC weight $w=0.996$. Second-best is the complicated version of the model without dependencies (green line, logprob $L=-396$, and BIC weight $w=0.004$). Third-best is our one-parameter model assuming independence (red line, $L=-419$, $w=3\cdot 10^{-11}$). And last (but not far from the third one) the model from Ref. \cite{Ward2008} (blue line, $L=-411$ $w=5\cdot 10^{-13}$). For the model from Ref. \cite{Ward2008}, $L$ and $w$ correspond to a re-optimization of the model as described in \meth, because using the parameters reported in \cite{Ward2008} would perform worse). 
}
\label{fig:S4}
\end{figure}

\begin{figure}[!ht]
\begin{center}
\includegraphics[width=0.4\columnwidth]{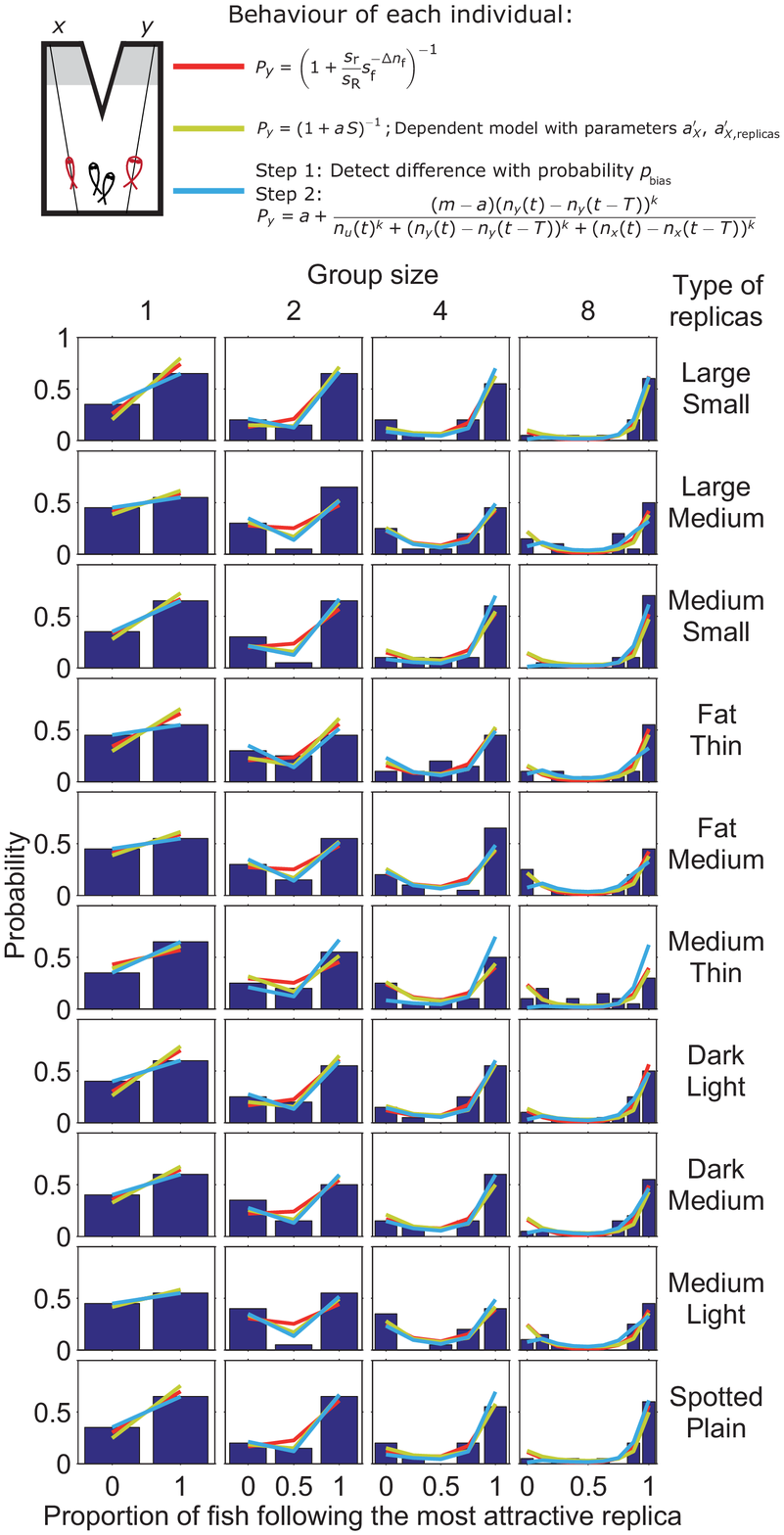}
\end{center}
\caption{
{\bf Comparison between different models for the condition with two different replicas.} Experimentally measured statistics of final configurations of fish choices from 20 experimental repetitions \cite{Sumpter2008} (blue histograms). Red line: results from model in \refeq{4} in the main text ($\s_\text{f}=2.9$, $\s_\text{r}/\s_\text{R}$ = 0.35, 0.7, 0.5, 0.52, 0.69, 0.75, 0.43, 0.55, 0.78, 0.43 for each row from top to bottom\comment{; red region: 95\% confidence interval}). Yellow line: Model including dependencies ($\ac_X'=4.8$, $\ac_{\X,\text{replicas}}'$ = 21.4, 11.8, 0.6, 9.9, 4.8, 0.9, 13, 8, 0.7, 14.5, 0.9 for each type of replica (large, medium, small, etc.).  Blue line: Empirical model presented in Ref. \cite{Sumpter2008}, using the parameters reported there. Different graphs correspond to different stickleback group sizes and different types of replicas going to $\x$ and $\y$. According to Bayesian Information Criterion (BIC, see \meth), our model neglecting dependencies gives the best representation of the data (red line, logprob $L=-783$, and BIC weight $w=0.9985$). Second-best is out model including dependencies, ($L=-788$, $w=0.001$). Last, but near the second one, is the model from ref. \cite{Sumpter2008} (blue line, $L=-781$ $w=0.0005$. For the model from Ref. \cite{Sumpter2008}, these values of $L$ and $w$ correspond to a re-optimization of the model as described in \meth, because using the parameters reported in \cite{Sumpter2008} would perform worse). The values of logprob ($L$) reported here do not include the data of the single-individual experiments (see \meth).
}
\label{fig:S5}
\end{figure}

\begin{figure}[!ht]
\begin{center}
\includegraphics[width=0.4\columnwidth]{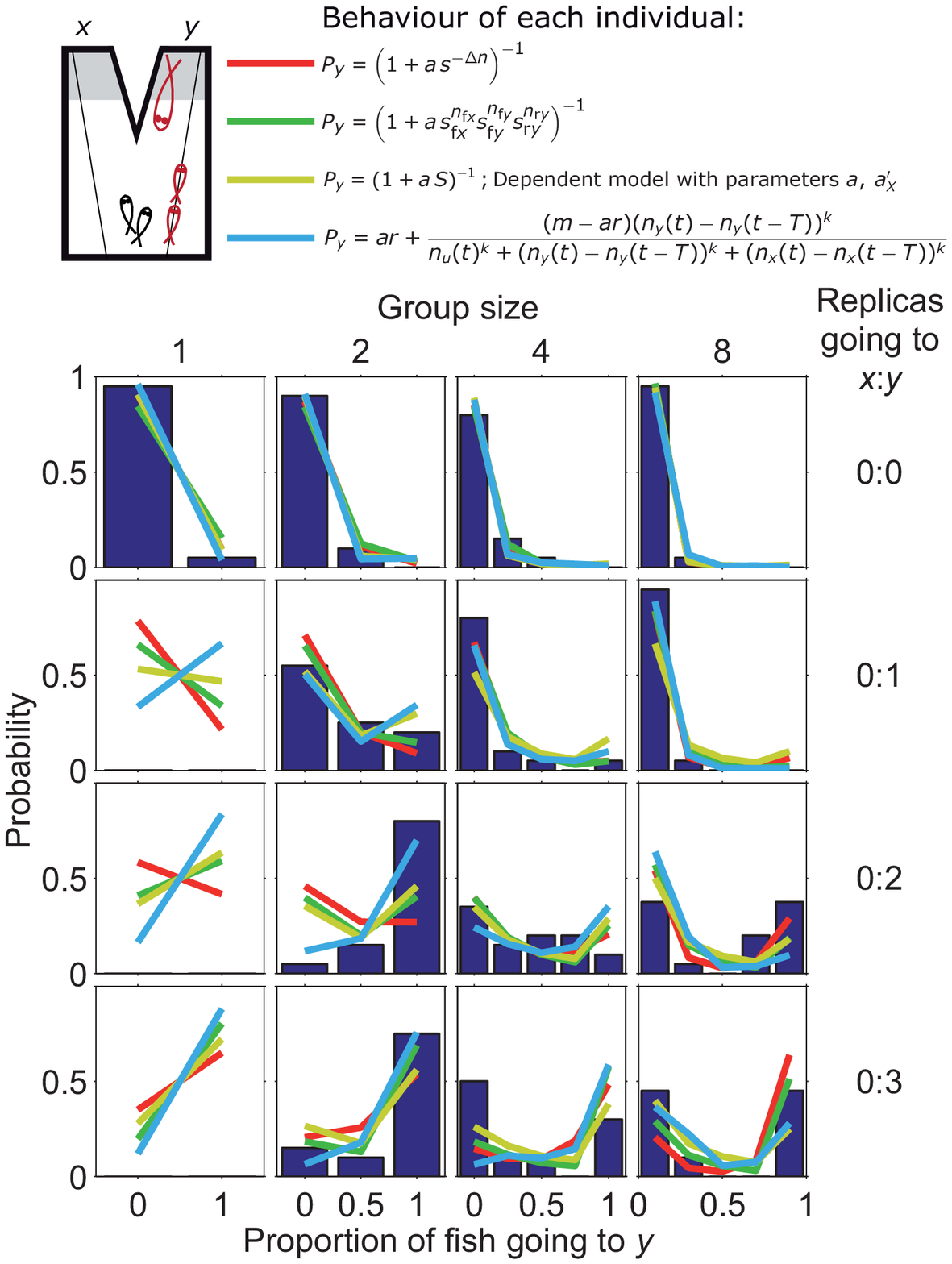}
\end{center}
\caption{
{\bf Comparison between different models in the asymmetrical set-up.} Experimentally measured statistics of final configurations of fish choices from 20 experimental repetitions \cite{Ward2008} (blue histograms). Red line: results from model neglecting dependencies in \refeq{5} in the main text ($\s=2.6$, $\ac=9.5$). Green line: Enhanced model neglecting dependencies with different reliability for the fish going to different locations and for the replicas ($\ac=5.5$, $\s_{\text{f}\x}=50$, $\s_{\text{f}\y}=2/3$, $s_{\text{r}\y}=0.36$. $\s_{\text{r}\x}$ has no effect because there are no replicas going to $\x$). Yellow line: Two-parameter model including dependencies ($\ac=9.94$, $\ac_\X'=8.66$). Blue line: Empirical model presented in Ref. \cite{Ward2008}, using the parameters reported there. Different graphs correspond to different stickleback group sizes and different number of replicas going to $\y$. According to Bayesian Information Criterion (BIC, see \meth), the best two models are our complicated version neglecting dependencies (green line, logprob $L=-225$, and BIC weight $w=0.52$) and our two-parameter model including dependencies (yellow line, $L=-231$, $w=0.38$). Next (but very near) is our simplified model (red line, $L=-232$, $w=0.098$). And last (and significantly worse) the model from Ref. \cite{Ward2008} (blue line, $L=-234$ $w=2.5\cdot 10^{-6}$. For the model from Ref. \cite{Ward2008}, the values of $L$ and $w$ correspond to a re-optimization of the model as described in \meth, because using the parameters reported in \cite{Ward2008} would perform worse. In two of the graphs for group size 1 that there are no data the prediction of the model from Ref. \cite{Ward2008} and our model (especially the simplest version) are opposite. It might be that the results changed completely, depending on the results of these graphs, were the experiments performed. But we found that this is not the case: We performed simulations, adding experimental data in these two graphs. Even in the extreme case that the fabricated results matched exactly the predictions of the model in Ref. \cite{Ward2008}, BIC would still favour two of our models (we would get $L=-254$, $w=0.99$ for our model with dependence, $L=-252$, $w=0.01$ for our complicated model neglecting dependence, $L=-268$, $w=8\cdot 10^{-7}$ for our simplified model neglecting dependence and $L=-258$, $w=3\cdot 10^{-6}$ for the model in \cite{Ward2008}). 
}
\label{fig:S6}
\end{figure}

\begin{figure}[!ht]
\begin{center}
\includegraphics[width=0.4\columnwidth]{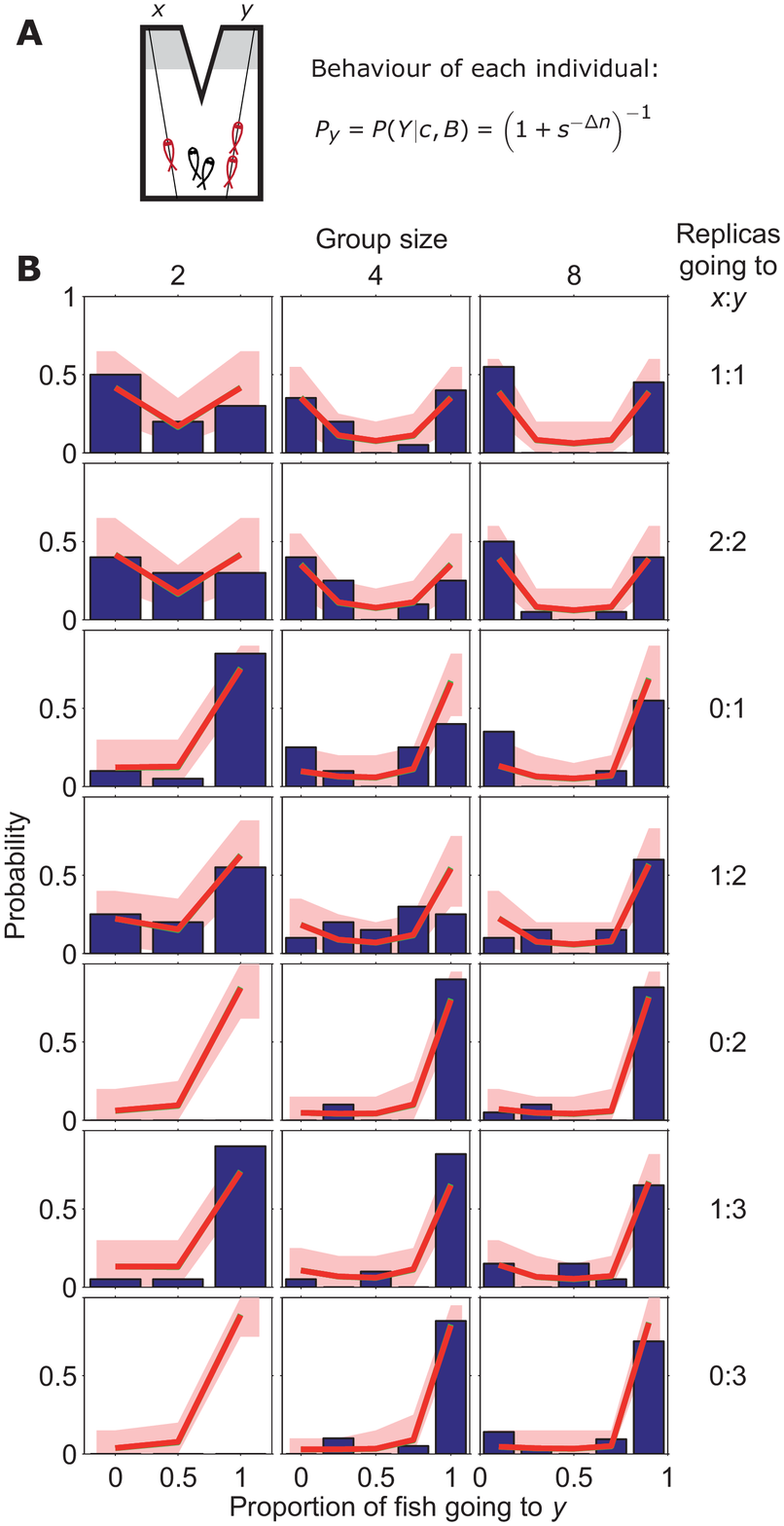}
\end{center}
\caption{
{\bf Comparison between model including dependencies and stickleback choices in symmetric set-up.}  (\emph{A}) Schematic diagram of symmetric set-up with a group of sticklebacks (in black) choosing between two identical refugia and with different numbers of replica fish (in red) going to $\x$ and $\y$. (\emph{B}) Experimentally measured statistics of final configurations of fish choices from 20 experimental repetitions \cite{Ward2008} (blue histogram) and results from the model that takes into account dependencies (red line using $\ac_\X'=4.9$; red region: 95\% confidence interval; green line with $\ac_\X'=5$). Different graphs correspond to different stickleback group sizes and different number of replicas going to $\x$ and $\y$.
}
\label{fig:datoswardnopredcaminos}
\end{figure}

\begin{figure}[!ht]
\begin{center}
\includegraphics[width=0.4\columnwidth]{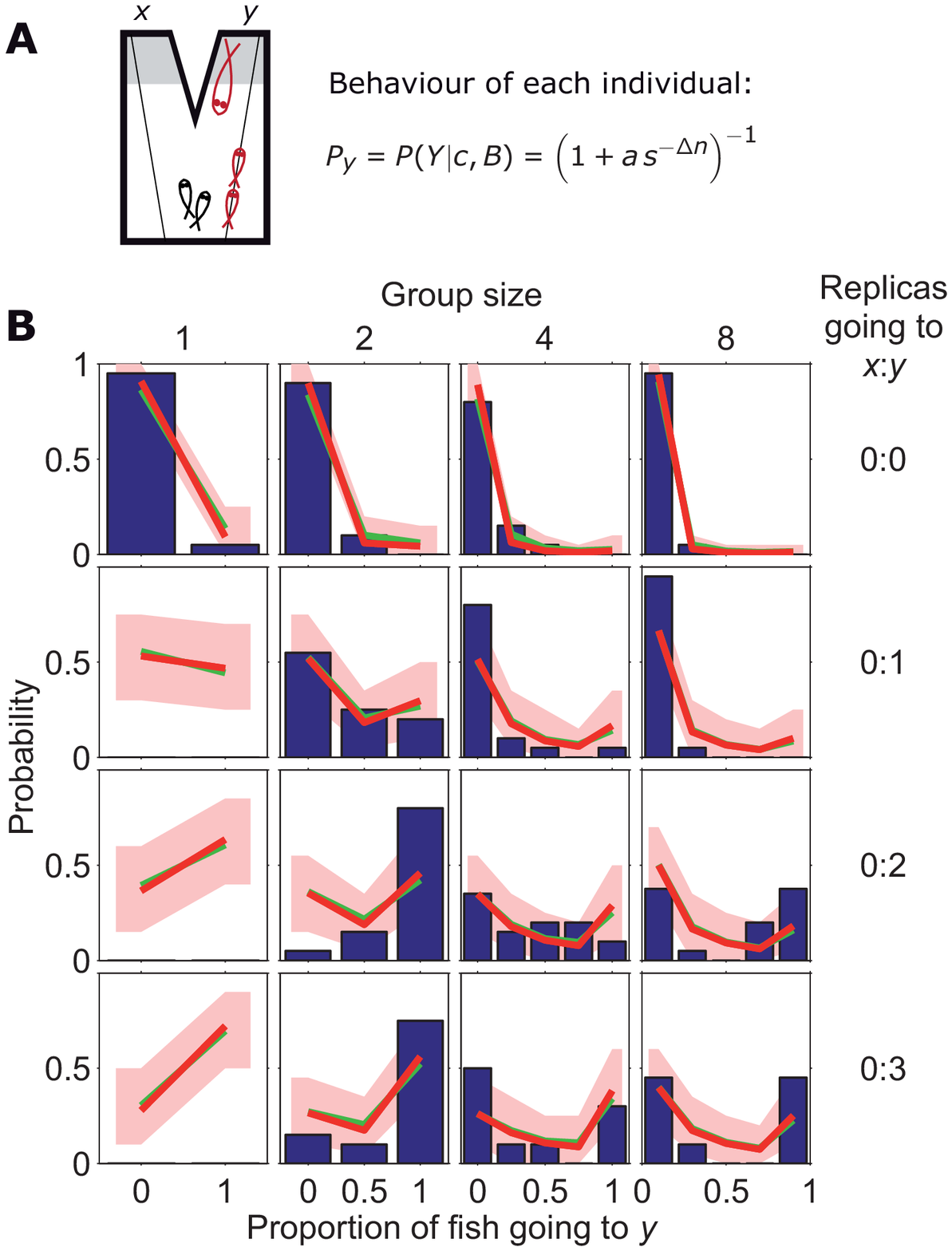}
\end{center}
\caption{
{\bf Comparison between model including dependencies and stickleback choices in asymmetric set-up.}  \emph{A}) Schematic diagram of asymmetric set-up (predator at $\y$, large fish depicted in red) with a group of sticklebacks (in black) choosing between two refugia, and replica fish (small fish depicted in red) going to $\y$. (\emph{B}) Experimentally measured statistics of final configurations of fish choices from 20 experimental repetitions \cite{Ward2008} (blue histogram) and results from the model that takes into account the dependencies (red line using $\ac_\X'=8.7$, $\ac=9.9$; red region: $95\%$ confidence interval. Green line using $\ac_\X'=5$ and $\ac=6.28$). Different graphs correspond to different stickleback group sizes and different number of replicas going to $\y$.
}
\label{fig:datoswardpredcaminos}
\end{figure}

\begin{figure}[!ht]
\begin{center}
\includegraphics[width=0.4\columnwidth]{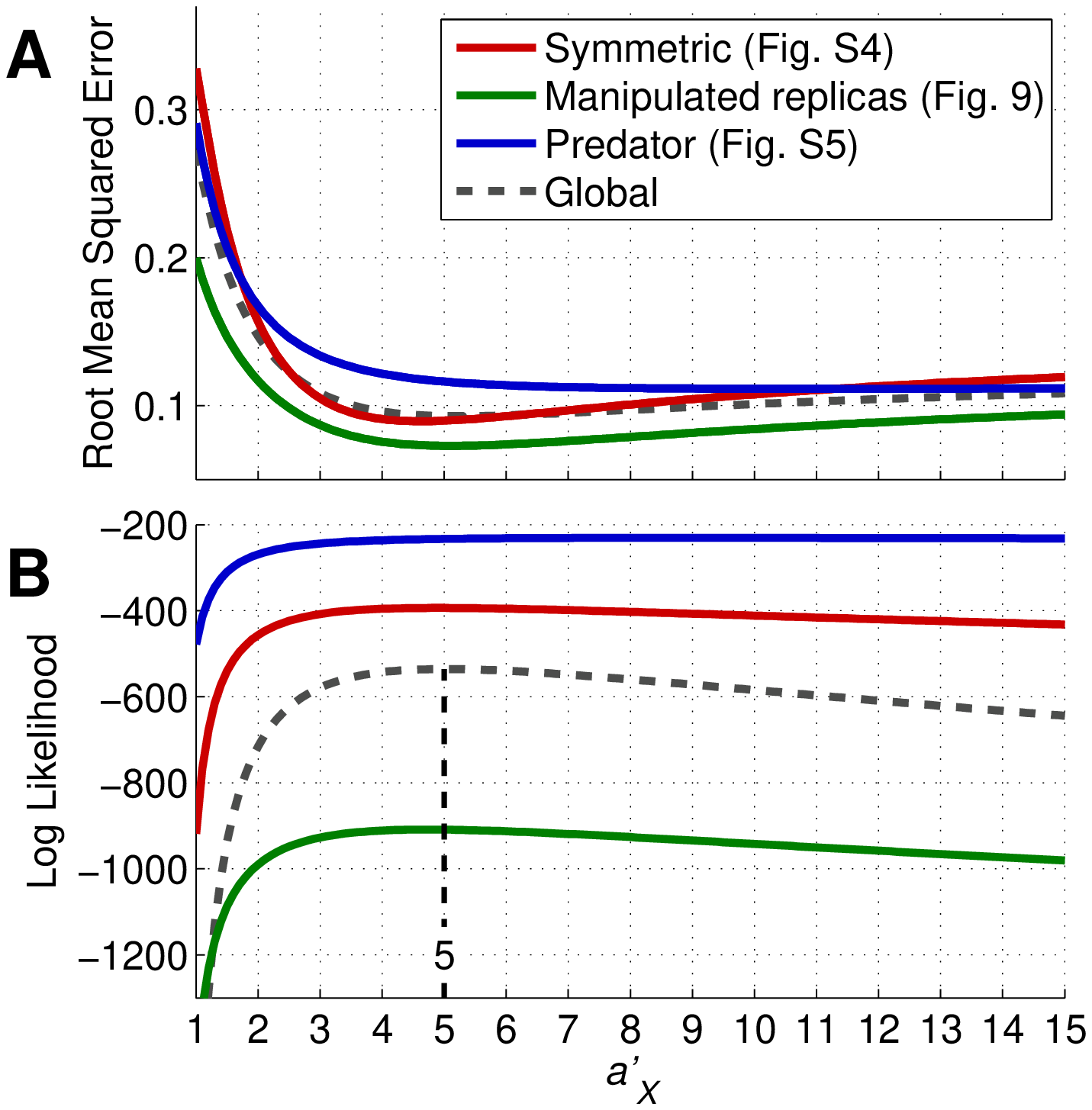}
\end{center}
\caption{
{\bf Goodness of fit of the model including dependencies for different values of $\ac_\X'$.}  \textbf{Red:} Symmetric case (data in \refig{datoswardnopredcaminos}{}). \textbf{Green:} Case with different replicas at each side (data in \refig{datosCurrBiolcaminos}{}. The parameters $\ac_{\X,\text{replicas}}'$ are re-optimized for each value of $\ac_\X'$). \textbf{Blue:} Asymmetric set-up with predator on one side (data in \refig{datoswardpredcaminos}{}; Parameter $\ac$ is re-optimized for each value of $\ac_\X'$). (\emph{A}) Root mean squared error between the data and the probabilities predicted by the model. Grey dashed line shows the mean RMSE for the three cases. The absolute values for each case depend on the shape of the data and are not comparable, only the trends and the position of the minima should be compared. (\emph{B}) Logarithm of the probability that the data come from the model. The height of each curve depends on the number of data for each experiment, only the trend and the position of the maxima should be compared. Grey dashed line shows the sum of the three coloured lines, but shifted by 1000 so that it fits on the scale. The peak of this global probability indicates the value of $\ac_\X'$ that best fits the three datasets ($\ac_\X'=5$).
}
\label{fig:fita1_caminos}
\end{figure}

\clearpage



\twocolumn
\renewcommand{\theequation}{S\arabic{equation}}
\setcounter{equation}{0}
\section*{Supporting text: Model for more than 2 options}

We present a derivation of the model for the more general case of \emph{M} different options (instead of the 2 options used in the main text). We also discuss some particular cases that give simple expressions while still widely applicable.


\subsection*{Model for \emph{M} options}
\fontfamily{cmr}\selectfont
Let $\M$ be the number of possible options, $\y_m,\ m=1\dots \M$.  Each individual estimates the probability that each option is the best one, using its non-social information ($\ns$) and the behavior of the other individuals ($\B$). So for one given option, say $\y_\tref$, we want to compute
\begin{equation}\label{ecn:S1}
P(\Y_\tref|\ns,\B),
\end{equation}
where $\Y_\tref$ stands for '$\y_\tref$ is the best option'. We can compute the probability in \refeq{S1} using Bayes' theorem,
\begin{equation}\label{ecn:Sbayes}
P(\Y_\tref|\ns,\B)=\frac{P(\B|\Y_\tref,\ns)P(\Y_\tref|\ns)}{\sum_{m=1}^M P(\B|\Y_m,\ns)P(\Y_m|\ns)}.
\end{equation}
Dividing numerator and denominator by the numerator, we get
\begin{equation}\label{ecn:general}
P(\Y_\tref|\ns,\B)=\frac{1}{\sum_{m=1}^M \ac_{m\tref} S_{m\tref}},
\end{equation}
where 
\begin{equation}
\ac_{m\tref}=\frac{P(\Y_m|\ns)}{P(\Y_\tref|\ns)}
\end{equation}
contains only non-social information, and
\begin{equation}\label{ecn:Ssup}
S_{m\tref}=\frac{P(B|\Y_m,\ns)}{P(B|\Y_\tref,\ns)}
\end{equation}
contains the social information. Note that each term of the summation preserves the multiplicative relation between social and non-social information that was also apparent in Eq. \ecngeneral\ of the main text. There may be $M-1$ independent non-social parameters $\ac_{m\tref}$ in the case that no two options have equal non-social information. But usually this will not be the case, and the number of independent non-social parameters will be lower.

Now we assume independence among behaviors (Eq. \ecnindep\ in main text), and group all possible behaviors in $L$ classes, $\{\beh{k}\}_{k=1}^L$ (Eq. \ecnkbehav\ in main text). These two assumptions transform \refeq{Ssup} into
\begin{equation}\label{ecn:Sindepsup}
S_{m\tref}=\prod_{k=1}^L \s_{k,m\tref}^{n_k},
\end{equation}
where $n_k$ is the number of individuals performing behavior $\beh{k}$, and
\begin{equation}\label{ecn:ak}
\s_{k,m\tref}=\frac{P(\beh{k}|\Y_m,\ns)}{P(\beh{k}|\Y_\tref,\ns)}
\end{equation}
are the reliability parameters for behavior $\beh{k}$ with respect to options $\y_m$ and $\y_\tref$. There may be up to $L(M-1)$ independent reliability parameters but usually they will not be all independent.

In summary, from Equations \textbf{\ref{ecn:general}} and \textbf{\ref{ecn:Sindepsup}} we have that
\begin{equation}\label{ecn:generalfinal}
P(Y_\tref|\ns,B)=\left(\sum_{m=1}^M{ \ac_{m\tref} \prod_{k=1}^L{\s_{k,m\tref}^{n_k}}}\right)^{-1}.
\end{equation}

This equation summarizes the general model applicable to any kind of experiment. In the following sections we consider two particular cases with a much simpler expression.

\subsection*{One basic reliability parameter}
The general model in \refeq{generalfinal} depends in general on $L(M-1)$ independent reliability parameters $\s_{k,m\tref}$. Here we derive the model for a particular case in which there is only one reliability parameter, $\s$. 

First, we consider classes of behaviors (from now on we call them just `behaviors') that simply consist of choosing a given option. If for example the options are different places, behaviors would be going to each of those places. Therefore, the number of possible behaviors is the same as the number of options, $L=M$. We use the convention that $\beh{j}$ is `choosing option $\y_j$'. Note that when a behavior is not informative (i.e. its reliability parameter is 1) it has no impact on the model in \refeq{generalfinal}. Therefore, considering this set of behaviors is equivalent to assuming that all other behaviors have reliability parameter equal to 1.

We further assume that $P(\beh{k}|\Y_m,\ns)$ only depends on whether $k=m$ or $k\neq m$, so that
\begin{equation}\label{ecn:probsiguales}
\begin{aligned}
&P(\beh{k}|Y_k,\ns)=P(\beh{l}|Y_l,\ns)\quad\\
&P(\beh{k}|Y_m,\ns)=P(\beh{l}|Y_p,\ns),\quad k\neq m,\ l\neq p
\end{aligned}
\end{equation}
Note that $P(\beh{k}|\Y_k,\ns)$ is the probability that another individual makes the correct choice, and $P(\beh{k}|\Y_m,\ns) with\ k\neq m$ is the probability that it makes a wrong choice. So this assumption means that the probability of making the correct choice is the same regardless of which option is actually the correct one. In the case of symmetric choices, in which non-social information $\ns$ is the same for all options, this relation will hold automatically, not being an extra assumption. It is likely that it also holds for many asymmetric choices. For example, the results for the asymmetric set-up presented in the main text suggest that it holds in that case. We define
\begin{equation}
\begin{aligned}
&p_c\equiv P(\beh{k}|Y_k,\ns)\\
&p_f\equiv P(\beh{k}|Y_m,\ns),\quad k\neq m.
\end{aligned}
\end{equation}
As it only matters whether the behavior matches the correct choice or not, there are only four distinct types of reliability parameters $\s_{k,m\tref}$ (\refeq{ak}):
\begin{equation}\label{ecn:tiposak}
\begin{aligned}
&\s_{k,kk}=\frac{P(\beh{k}|\Y_k,\ns)}{P(\beh{k}|\Y_k,\ns)}=\frac{p_c}{p_c}=1\\
&\s_{k,ml}=\frac{P(\beh{k}|\Y_m,\ns)}{P(\beh{k}|\Y_l,\ns)}=\frac{p_f}{p_f}=1,\quad k\neq m,\ k\neq l\\
&\s_{k,km}=\frac{P(\beh{k}|\Y_k,\ns)}{P(\beh{k}|\Y_m,\ns)}=\frac{p_c}{p_f}=\s,\quad k\neq m\\
&\s_{k,mk}=\frac{P(\beh{k}|\Y_m,\ns)}{P(\beh{k}|\Y_k,\ns)}=\frac{p_f}{p_c}=\frac{1}{\s},\quad k\neq m,\\
\end{aligned}
\end{equation}
where
\begin{equation}\label{ecn:a}
\s\equiv \frac{p_c}{p_f}
\end{equation}
is the basic reliability parameter, equal to the probability that another individual makes the correct choice over the probability that it makes a mistake, for any behavior and for any individual. We regroup the terms in\refeq{generalfinal} so that it reflects the different types of $\s_{k,m\tref}$ (\refeq{tiposak}), and get
\begin{multline}\label{ecn:generalreordenada}
P(Y_\tref|\ns,B)=\\\Bigg(\sum_{m=1}^M{ \ac_{m\tref} \s_{m,m\tref}^{n_m} \s_{\tref,m\tref}^{n_\tref} \prod_{\substack{k=1\\k\neq m\\ k\neq\tref}}^L{\s_{k,m\tref}^{n_k}}}\Bigg)^{-1}.
\end{multline}
Using the relations in \refeq{tiposak} we have that
\begin{equation}\label{ecn:generalacomun}
P(Y_\tref|\ns,B)=\left(\sum_{m=1}^M{ \ac_{m\tref} \s^{-(n_\tref-n_m)}}\right)^{-1}.
\end{equation}
Note that the term $m=\tref$ is always equal to 1, so \refeq{generalacomun} is identical to
\begin{equation}\label{ecn:generalacomun1mas}
P(Y_\tref|\ns,B)=\Bigg(1+\sum_{\substack{m=1\\m\neq \tref}}^M{ \ac_{m\tref} \s^{-(n_\tref-n_m)}}\Bigg)^{-1},
\end{equation}
that has the same structure as the equations presented in the main text.

\subsection*{Symmetric case}
In the special case that all options are indistinguishable using non-social information alone (symmetric case), all non-social parameters $\ac_{m\tref}$ are equal 1 and \refeq{generalacomun1mas} becomes
\begin{equation}\label{ecn:generalsim}
P(\Y_\tref|\ns,\B)=\Bigg(1+\sum_{\substack{m=1\\m\neq\tref}}^M \s^{-(n_\tref-n_m)}\Bigg)^{-1}.
\end{equation}
We recall that in this case \refeq{probsiguales} holds automatically, not being an extra assumption.

\begin{figure}
\begin{center}
\includegraphics[width=\columnwidth]{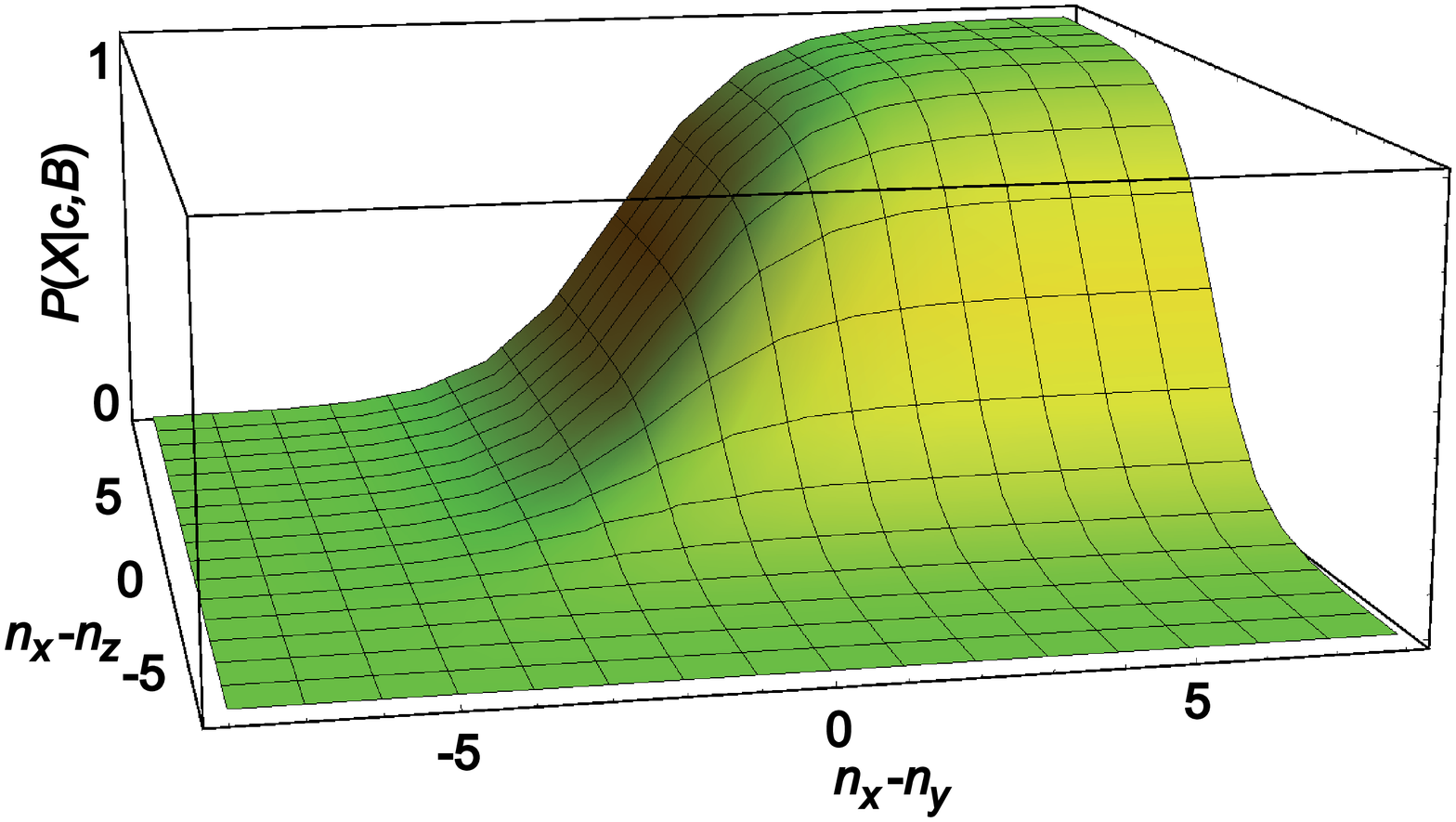}
\end{center}
\caption{
Probability of choosing one of the options for the 3-choice symmetric case. 
}
\label{fig:S5S}
\end{figure}

In the particular case of 3 options, $x$, $y$, $z$, we have
\begin{equation}\label{ecn:3options}
P(X|\ns,\B)=\left(1+\s^{-(n_x-n_y)}+\s^{-(n_x-n_z)}\right)^{-1},
\end{equation}
and the corresponding expressions for $P(Y|\ns,\B)$ and $P(Z|\ns,\B)$. \refig{S5S}{} shows $P(X|\ns,\B)$ in terms of its two effective variables, $n_x-n_y$ and $n_x-n_z$ (\refeq{3options}).

\end{document}